\documentclass[aps,pra,twocolumn,superscriptaddress]{revtex4-2}
\usepackage{graphicx,dcolumn,hyperref,braket,amsmath,amssymb,xcolor} 
\usepackage{epstopdf}
\usepackage{natbib}
\usepackage{mathtools}
\usepackage{multirow}
\usepackage{xcolor}
\newcommand{\cor}[1]{#1}

\begin{document}

\title{\cor{Testing higher-order quantum interference with many-particle states}}

\author{Marc-Oliver Pleinert}
\affiliation{Institut f\"{u}r Optik, Information und Photonik, \\ Friedrich-Alexander-Universit\"{a}t Erlangen-N\"{u}rnberg (FAU), 91058 Erlangen, Germany}
%\affiliation{International Max Planck Research School - Physics of Light (IMPRS-PL), Max Planck Institute for the Science of Light, 91058 Erlangen, Germany}
\affiliation{Erlangen Graduate School in Advanced Optical Technologies (SAOT), Friedrich-Alexander-Universit\"{a}t Erlangen-N\"{u}rnberg (FAU), 91052 Erlangen, Germany}

\author{Alfredo Rueda}
\affiliation{Institut f\"{u}r Optik, Information und Photonik, \\ Friedrich-Alexander-Universit\"{a}t Erlangen-N\"{u}rnberg (FAU), 91058 Erlangen, Germany}
\affiliation{{\rm{currently at:}} Scantinel Photonics, Carl-Zeiss-Strasse 22, 73447 Oberkochen, Germany}

\author{Eric Lutz}
\affiliation{Institute for Theoretical Physics I, University of Stuttgart, D-70550 Stuttgart, Germany}

\author{Joachim von Zanthier}
\affiliation{Institut f\"{u}r Optik, Information und Photonik, \\ Friedrich-Alexander-Universit\"{a}t Erlangen-N\"{u}rnberg (FAU), 91058 Erlangen, Germany}
\affiliation{Erlangen Graduate School in Advanced Optical Technologies (SAOT), Friedrich-Alexander-Universit\"{a}t Erlangen-N\"{u}rnberg (FAU), 91052 Erlangen, Germany}

\begin{abstract}
Quantum theory permits interference between indistinguishable paths but, at the same time, restricts its order. 
Single-particle \cor{interference}, for instance, is limited to the second order, that is, to pairs of single-particle paths. 
To date, all experimental efforts to search for higher-order interferences beyond those compatible with quantum mechanics have been based on such single-particle schemes. 
However, quantum physics is not bounded to single-particle \cor{interference}. 
We here experimentally study  many-particle higher-order \cor{interference} using a two-photon-five-slit setup. 
We observe nonzero two-particle \cor{interference} up to fourth order, corresponding to the interference of two distinct two-particle paths. 
We further show that fifth-order interference is restricted to $10^{-3}$ in the intensity-correlation regime and to $10^{-2}$ in the photon-correlation regime, thus providing novel bounds on \cor{higher-order quantum interference}.
\end{abstract}

\maketitle

%\section{Introduction}

Interference phenomena for both light and matter are an intrinsic property of quantum physics. They occur when indistinguishable paths exist as in Young's paradigmatic double-slit experiment~\cite{fey65}. 
From a fundamental point of view, quantum interference stems from coherent superpositions of states and thus from the linearity of quantum theory~\cite{per02}. 
However, quantum mechanics not only enables but also restricts interference~\cite{Sorkin:1994}. 
For instance, according to Born's rule, which relates detection probabilities to the modulus square of the wave function~\cite{Born:1926}, single-particle interference is limited to the second order, that is, to two interfering single-particle paths. 
In a multi-slit setup, interference is therefore expected to occur only between pairs of indistinguishable paths, and all higher orders in the interference hierarchy vanish~\cite{Sorkin:1994}.  

The physical origin of the lack of higher-order quantum interferences is not yet understood~\cite{Sorkin:1994}.
Their existence would have profound implications for quantum theory, including nonlocality and contextuality~\cite{Niestegge:2013,Dakic:2014,Henson:2015,Zhao:2018}. 
They have indeed been linked to violations of the spatial~\cite{Niestegge:2013} and temporal~\cite{Dakic:2014} Tsirelson bounds, as well as to a weakening of noncontextuality bounds~\cite{Henson:2015}. 
They would hence reveal entanglement stronger than predicted by quantum theory. 
They would further permit perfect interaction-free measurements~\cite{Zhao:2018}. 
For this reason, a growing number of single-particle experiments have been realized in the past years to detect such higher-order interference, using photons in the optical~\cite{Sinha:2010,Hickmann:2011,Sollner:2012,Kauten:2017} and microwave~\cite{Rengaraj:2018} domain, as well as molecules~\cite{Cotter:2017}, atoms~\cite{Barnea:2018} and spin systems~\cite{Park:2012,Jin:2017}. \cor{However, while high-sensitivity tests of the linearity of quantum mechanics have been performed~\cite{Shull:1980,Gahler:1981,Bollinger:1989,Chupp:1990,Walsworth:1990,Majumder:1990,Vinante:2017}, similar experiments on higher-order interference are missing.}

Quantum physics goes beyond single-particle interference by allowing for {many-particle} interference in the case of indistinguishable particles. 
A prominent example is provided by the Hong-Ou-Mandel experiment, in which two non-interacting photons can influence each other via two-particle interference~\cite{Hong:1987}. 
Many-particle interference, \cor{described by Glauber's theory of quantum optical coherence~\cite{Glauber:1963}}, is mathematically richer and physically more subtle than single-particle interference~\cite{Pan:2012,Tichy:2014,Agne:2017,Menssen:2017}. It has found  widespread applications in metrology~\cite{Giovannetti:2011,Su:2017}, imaging~\cite{Hanbury-Brown:1956,Thiel:2007,Schneider:2018a}, and quantum information processing~\cite{Aaronson:2011,Wang:2019}. 
Recently, the interference hierarchy has been theoretically extended to the general case of $M$-particle interference with $N$ modes~\cite{Pleinert:2020}. 
In this situation, Born's rule allows for higher-order  path interference of up to order $2M$. 
In addition, owing to the much larger number of interfering paths, many-particle interference has been predicted to offer increased sensitivity to deviations from quantum theory compared to its single-particle counterpart~\cite{Pleinert:2020}. 

We here report the first experimental investigation of many-particle higher-order quantum \cor{interference} using a two-photon-five-slit setup.
We determine single-particle and two-particle \cor{interferences} up to  fifth order, both in the intensity and in the photon-counting regimes. To this end, we measure and evaluate first-order and second-order \cor{(spatial)} correlation functions for a total of $2^5=32$ different interference configurations. While single-particle \cor{interference} vanishes at the third order, we show  that two-particle \cor{interference} only cancels at the fifth order, in agreement with standard quantum theory.

\paragraph*{Experimental setup.}

\begin{figure}[t]
	\centering \includegraphics[width=\columnwidth]{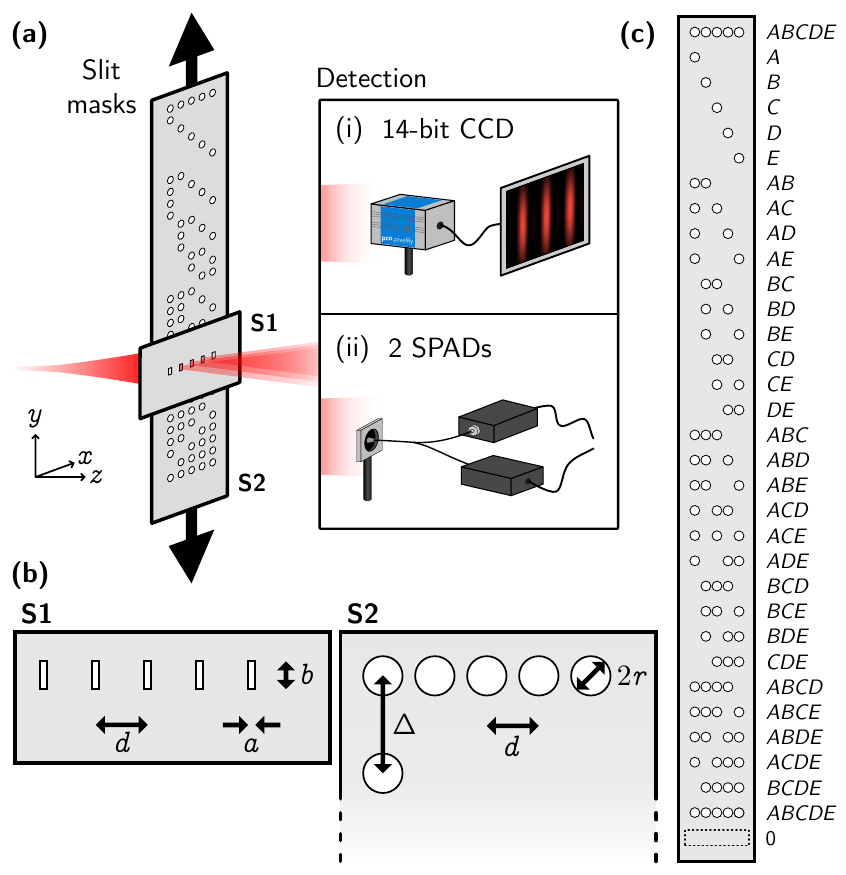}
	\caption{\label{fig:setup}(a) Setup to measure the interference hierarchy in single-particle and two-particle \cor{interferences}. Coherent photons are sent through two slit masks, S1 and S2, and measured in the far field, either (i) by a charge coupled device (CCD) or (ii) by two single photon avalanche diodes (SPADs). (b)~Fixed base slit mask S1 with dimensions $a=25\,\mu\text{m}$, $b=200\,\mu\text{m}$ and $d=500\,\mu\text{m}$. The dimensions of the movable slit mask S2 are $2r=400\,\mu\text{m}$ with a spacing $\Delta=1000\, \mu\text{m}$.  (c)~The layout of slit mask S2 contains $33$ configurations.}
\end{figure}

In order to assess single-particle and two-particle interference orders, we realize and analyze different experimental arrangements [Fig.~\ref{fig:setup}(a)]: 
Photons, in a coherent state $\ket{\alpha}$ with mean photon number $\bar{n}=|\alpha|^2$, are provided by a HeNe laser at $\lambda=633\,\text{nm}$.
These photons are scattered at two slit masks S1 and S2 [Figs.~\ref{fig:setup}(b-c)]. 
The base slit mask S1 is a five-fold slit (denoted $ABCDE$), while the movable blocking slit mask S2 consists of $33$ configurations. 
Both masks have the same distance of adjacent slits $d=500\,\mu\text{m}$.
By moving the blocking slit mask S2 in front of the fixed slit mask S1, we can implement all possible slit arrangements from one to five slits.
For instance, in Fig.~\ref{fig:setup}(a), slits $A$ and $E$ of S1 are blocked by S2 such that the three-slit configuration $BCD$ is realized. 
The measurement is conducted in the far field at a distance $L=1.7\,\text{m}$ behind the first slit mask 
either (i) in the intensity regime by a specialized, high-performance $14$-bit charge-coupled device (CCD) camera 
or (ii) in the photon-counting regime by two fiber-coupled single-photon avalanche diodes (SPADs), whose signals are time-registered and correlated~\cite{Supplemental}. In this far-field regime, effects coming from nonclassical looped paths \cite{Yabuki:1986,Sawant:2014,Magana-Loaiza:2016} are negligible \cite{Pleinert:2020}. 

During a measurement sequence, the source, the base mask S1 and the detection systems are fixed, while  the blocking mask S2 is scanned in a fully automatized way to reduce alignment errors~\cite{Supplemental}. Data acquisition is also fully automatized.
A second five-fold slit $ABCDE$ has moreover been added at the top of S2 to compare the interference patterns at the beginning and the end of mask S2 and thus verify the initial alignment of the setup \cite{Supplemental}.

The measurement results can be interpreted in the few-particle regime by expanding the coherent state as
\begin{equation}
\ket{\alpha} = c_0\ket{0}+c_1\ket{1}+c_2\ket{2}+\dots \, ,
\end{equation}
where $p_n=|c_n|^2$, $(n = 0,1, 2, \dots)$, is the Poissonian probability of the $n$-photon state to occur.
Whenever we register a photon at only one of the two SPADs, the effective state is given by $\ket{1}$, which has been coherently distributed over the slits, leading to single-particle interference. For a coincident detection at both detectors, the effective state is $\ket{2}$, yielding two-particle interference. 
Accordingly, we can measure both interference hierarchies simultaneously and differentiate between them by filtering the events via postselection~\footnote{Since the SPADs are not number-resolving, multi-photon events at a single SPAD cannot be excluded. However, such events have a negligible effect on our results  in the parameter regime of the experiment~\cite{Supplemental}.}.

\paragraph*{Interference hierarchy.}

Although they are noninteracting, identical photons can influence each other via interference of distinct but indistinguishable $M$-particle paths. 
Such $M$-particle paths lead to \cor{(spatial)} correlations between field modes that can be conveniently captured, on the photon as well as on the intensity level, by the $M$th-order  correlation function~\cite{Glauber:1963},
\begin{eqnarray}\label{eq:GM}
G^{(M)}(\delta_1,\ldots,\delta_M) \propto  \braket{\hat{a}^\dagger_1 \ldots \hat{a}^\dagger_M \hat{a}^{\phantom{\dagger}}_M \ldots \hat{a}^{\phantom{\dagger}}_1},
\end{eqnarray}
where $\hat{a}_i \equiv  \hat{a}(\delta_i)$ is the annihilation operator of the spatial mode $\delta_i=kd\sin \theta_i$, determined by the wave vector $k$, the slit distance $d$ and the angle of the $i$th detector $\theta_i$.
Interference in such $M$-particle correlations can be classified into various orders $I^{\small{(M)}}_N$, depending on how many different input modes ($A$, $B$, $\ldots$) interfere with each other.

\begin{figure*}
	\centering \includegraphics[width=\textwidth]{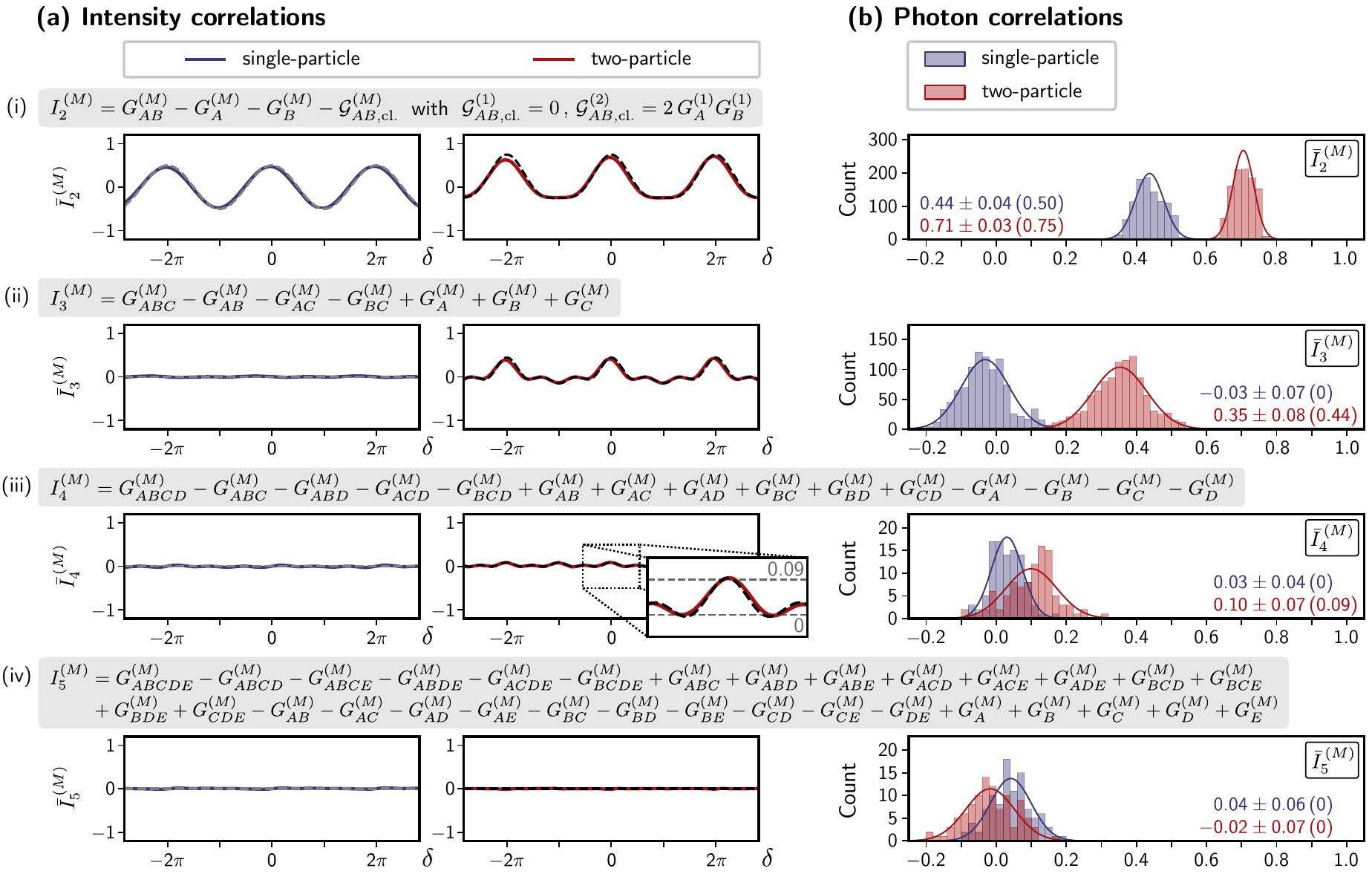}
	\caption{\label{fig:SP-vs-TP-hierarchy} Interference hierarchy in (a) the intensity-correlation and (b) the photon-correlation regime from (i) second order to (iv) fifth order (related formulas are taken from Ref.~\cite{Pleinert:2020}). (a)~Experimental data (solid) and quantum theory (dashed) of single-particle (blue, left) and two-particle (red, right) interference orders. Fourth-order two-particle interference exhibits a tiny modulation, clearly visible with a good signal-to-noise ratio and most pronounced at the center (inset). (b)~Histograms of the single-particle (blue) and two-particle (red) interference hierarchy at $\delta=0$ in the photon-counting regime.}
\end{figure*}

For single particles ($M=1$), the first-order interference is trivially given by the (relative) detection probability in the far field,
$I^{(1)}_1=P_A=G_A^{(1)}$, for a single slit $A$. The second-order interference is obtained by comparing the quantum-mechanical double-slit ($AB$) signal with the classical sum of the two single slits ($A$ and $B$)~\cite{Sorkin:1994},
\begin{eqnarray}
I^{(1)}_2=G^{(1)}_{AB}-(G^{(1)}_{A}+G^{(1)}_{B})\, .
\end{eqnarray}
Higher orders can be constructed accordingly~\cite{Sorkin:1994,Pleinert:2020} and the explicit expressions are given in Fig.~\ref{fig:SP-vs-TP-hierarchy}. 
For single-particle correlations, all higher-order terms vanish, $I^{(1)}_3=I^{(1)}_4= I^{(1)}_5= \ldots=0$~\cite{Sorkin:1994}. 
The single-particle interference hierarchy hence truncates at the third order.

By contrast, for two particles ($M=2$), nonzero interference occurs up to the fourth order \cite{Pleinert:2020},
\begin{align}
I^{(2)}_2 \neq 0, I^{(2)}_3 \neq 0, I^{(2)}_4 \neq 0 \text{ ; } I^{(2)}_5 = I^{(2)}_6=\ldots=0 \, .
\end{align}
Born's rule hence allows for the interference of two two-particle paths, and the two-particle interference hierarchy is only truncated at the fifth order \footnote{The two-particle Hong-Ou-Mandel experiment corresponds to second-order interference, since it only involves two input modes~\cite{Hong:1987}.}.

The vanishing of many-particle higher-order \cor{interference}  is captured by the $M$-particle Sorkin parameter defined as the normalized $(2M+1)$th interference order~\cite{Pleinert:2020},
\begin{eqnarray}\label{eq:Sorkin-family}
\kappa^{(M)}=\frac{I^{(M)}_{2M+1}}{G^{(M)}_{A, B, C, \ldots}(0, 0, 0, \ldots)} \, ,
\end{eqnarray}
where the first member ($M=1$) is the single-particle Sorkin parameter~\cite{Sorkin:1994}. According to quantum mechanics, Eq.~\eqref{eq:Sorkin-family} is zero for all $M$.

\paragraph*{Experimental results.}
To obtain the complete interference hierarchy of single-particle and two-particle correlations, we perform $31+1+1$ (correlation) measurements. The latter consist of the measurement of $31$ different slit configurations needed to evaluate the interference orders, one additional measurement for the second $ABCDE$ arrangement added at the top of the slit mask S2, and one final measurement of the background ($0$), where the mask S2 blocks all slits of the base mask S1. 
These $33$ measurements form a measurement set.

At the beginning of such a set, the measurement sequence is randomized to reduce systematic errors. 
A motorized translation stage addresses the slit mask S2 and implements the drawn slit configuration $X\in\{{ABCDE}(1), {A}, {B}, \ldots, {ABCDE}(2),0\}$. 
In the intensity regime, we take $250$ CCD images of each slit configuration.
The integration time $t_i = 2\,\text{ms}$ is fixed for all configurations and fully covers the CCD's dynamical range when measuring $ABCDE$.
In the photon-counting regime, the two SPADs register the single-photon and two-photon events within a fixed total time of $T=120~\rm{s}$ per slit configuration. 
Time tags of the photon events are registered by a time-to-digital converter and correlated within a time frame of $t_f=1\,\text{ns}$.
We ensure that the count rates are  $ \lesssim 100\,\rm{kHz}$ such that detector nonlinearities can be neglected~\cite{Supplemental}. 
We record the data in the autocorrelation scheme, where both detectors are effectively at the same position ($\delta_1=\delta_2=\delta$). 
This scheme is least sensitive to alignment errors. 
For intensity measurements, 
each pixel of the CCD can be regarded as an independent detector, and the autocorrelation function is measured by correlating pixels of the same optical phase $\delta$ from neighboring lines of the CCD~\cite{Supplemental}. 
On the other hand, for photon-counting measurements, the autocorrelation is implemented by a fiber beam splitter at $\delta$, connected to the two different SPADs.

In total, we perform $100$ such measurement sets, each with a different sequence, in the intensity regime as well as in the photon-counting regime.
For each set, the data is averaged per slit configuration and corrected by subtracting background and detector noise. 
From the corrected data, we evaluate the two-particle interference orders, $I^{(2)}_2,I^{(2)}_3,I^{(2)}_4, I^{(2)}_5$ as well as the single-particle interference orders, $I^{(1)}_2,I^{(1)}_3,I^{(1)}_4,I^{(1)}_5$, from (a subset of) the data as indicated in Fig.~\ref{fig:SP-vs-TP-hierarchy}.
We normalize the interference orders by the central value of the configuration with the most slits within a given order to remove any experiment-specific proportionality factors~\cite{Magana-Loaiza:2016,Cotter:2017,Barnea:2018}.
For example, for two slits $A$ and $B$, we use $\bar{I}^{\,(2)}_{2,AB}=I^{(2)}_{2,AB}/G^{(2)}_{AB}(0,0)$, where we have explicitly indicated the involved slits. 
For the fifth order in two-particle correlations, this corresponds to the normalized two-particle Sorkin parameter of Eq.~\eqref{eq:Sorkin-family}, $\kappa^{(2)} \equiv \bar{I}^{\,(2)}_{5}$.

In the intensity regime, the CCD covers the interval $\delta \in \left[ - 3 \pi, + 3 \pi \right]$ of the interference pattern and thus reveals the spatial behavior of the interference orders. 
The results for single-particle and two-particle correlations are shown in Fig.~\ref{fig:SP-vs-TP-hierarchy}(a): Solid lines correspond to experimental data and dashed lines to  quantum  theory. 
Single-particle interference (blue, left) vanishes starting with the third order, $\bar{I}^{\,(1)}_N=0$ for $N \geq 3$, with an uncertainty of $10^{-3}$, essentially limited by the relative misalignment of the slit configurations.
By contrast, two-particle interference $\bar{I}^{\,(2)}_N$ (red, right) also exhibits a nonzero third and fourth order, while only the fifth order disappears. 
The modulation of the fourth-order two-particle interference [inset of Fig.~\ref{fig:SP-vs-TP-hierarchy}(a)]
is of the order of $10^{-1}$, clearly identifiable with a good signal-to-noise ratio. %

\begin{figure}
	\centering \includegraphics[width=\columnwidth]{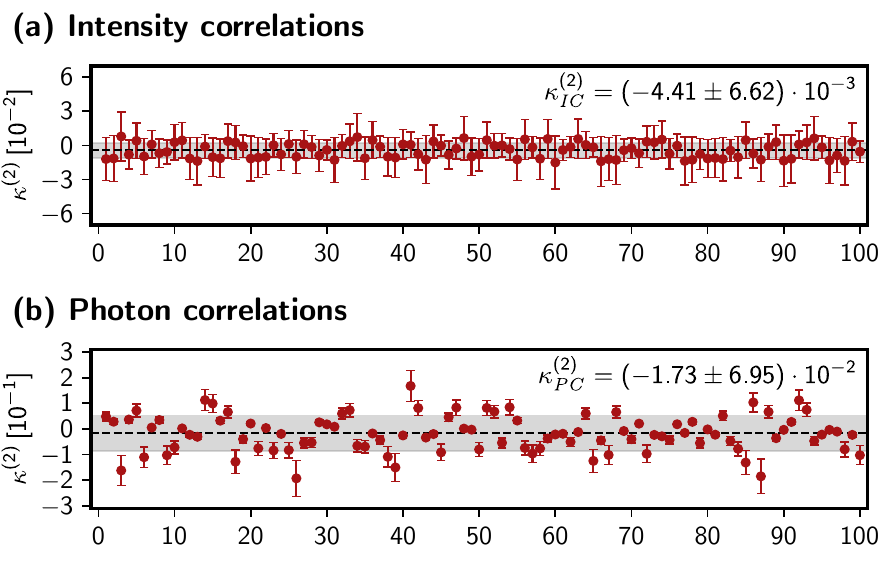}
	\caption{\label{fig:2P-Sorkin}Two-particle Sorkin parameter $\kappa^{(2)}$ from Eq.~\eqref{eq:Sorkin-family} for $100$ measurement sets in (a)~the intensity-correlation regime and (b)~the photon-correlation regime. In~(b), the error bars are enlarged by a factor $100$.}
\end{figure}

The difference between single-particle and two-particle correlations can be seen most prominently at $\delta=0$ also in the photon-counting regime [Fig.~\ref{fig:SP-vs-TP-hierarchy}(b)].
While single-particle and two-particle correlations exhibit both nonzero interference of the second order (though with a different value due to the normalization),
the difference between the two is clearly visible for the third-order term with
$\bar{I}^{\,(1)}_3=-0.03\pm0.07$ being effectively zero and  $\bar{I}^{\,(2)}_3=0.35\pm0.08$ being statistically different from zero~\footnote{Second- and third-order terms are a little smaller than expected, which suggests a slight off-center measurement.}.
The same holds true for the interference of the fourth order  with $\bar{I}^{\,(1)}_4=0.03\pm0.04$ and  $\bar{I}^{\,(2)}_4=0.10\pm0.07$~\footnote{Due to a difficult alignment, this result has been obtained with $100$ re-measurements of only the configuration $ABCD$ and its sub-configurations.}. 
On the other hand, the fifth-order interference is effectively zero for both single-particle and two-particle correlations.

The fifth-order term $\bar{I}^{\,(2)}_{5}$ can be used to rule out higher-order interference in two-particle correlations via the Sorkin parameter of Eq.~\eqref{eq:Sorkin-family}.
The experimental findings for $\kappa^{(2)}$ are shown in Fig.~\ref{fig:2P-Sorkin} for (a) intensity correlations (with statistical errors resulting from averaging over different pixels) and (b) photon correlations (with Poissonian errors), each consisting of $100$ independent sets of measurements. 
We obtain $\kappa^{(2)}_{IC}=(-4.41\pm 6.62)\cdot 10^{-3}$ in the intensity-correlation (IC) regime and $\kappa^{(2)}_{PC}=(-1.73\pm 6.95)\cdot 10^{-2}$ in the photon-correlation (PC) regime.

\paragraph*{Conclusions.}

We have performed a detailed experimental study of many-particle higher-order  interference  using a two-photon-five-slit setup, both in the intensity and in the photon-counting regimes. 
\cor{We have observed for the first time fourth-order two-particle interference, corresponding to the interference of two different two-particle paths employing four distinct modes~\cite{Note2}. We have, moreover, established the absence of the corresponding fifth-order interference at the level  of $10^{-2}-10^{-3}$}. 
\cor{The precision of our experiment may be improved further by reducing measurement errors using, for instance, a more stable integrated photonic network scheme with lower losses compared with our free-space setup~\cite{Keil:2016}. The present two-particle interference scheme leads to a tenfold increase in sensitivity~\footnote{\cor{The sensitivity quantifies the smallest change that can be detected in a measurement and depends on the measurement method. On the other hand, the precision is related to the measurement errors of the experimental setup.}} to deviations from quantum theory compared to existing single-particle experiments~\cite{Pleinert:2020}. Due to the exponential increase of interfering paths with growing particle number, this sensitivity  is expected to increase significantly: from one order of magnitude for the considered two-particle example up to 12 orders of magnitude for eight-particle interference~\cite{Pleinert:2020}.  
Since eight-particle experiments  have already been realized~\cite{Oppel:2014}, many-particle interference appears to be a promising approach for high-sensitivity tests of higher-order quantum interference.} 
An extension towards continuous variables might also be worthwhile for future investigations~\cite{Sabuncu:2007,Lassen:2010,Lassen:2010a}.

\paragraph*{Acknowledgments.}

M.-O.P. acknowledges support by the Studienstiftung des deutschen Volkes and the International Max Planck Research School - Physics of Light (IMPRS-PL). 
M.-O.P. and J.v.Z.  acknowledge funding by the Erlangen Graduate School in Advanced Optical Technologies (SAOT) by the German Research Foundation (DFG) in the framework of the German excellence initiative.
We thank Dr. Steffen Oppel for fruitful discussions at an early stage of the project and Dr. Irina Harder from the TDSU 1: Micro- \& Nanostructuring at the Max Planck Institute for the Science of Light for producing the slit mask S1.


\begin{thebibliography}{52}%
	\makeatletter
	\providecommand \@ifxundefined [1]{%
		\@ifx{#1\undefined}
	}%
	\providecommand \@ifnum [1]{%
		\ifnum #1\expandafter \@firstoftwo
		\else \expandafter \@secondoftwo
		\fi
	}%
	\providecommand \@ifx [1]{%
		\ifx #1\expandafter \@firstoftwo
		\else \expandafter \@secondoftwo
		\fi
	}%
	\providecommand \natexlab [1]{#1}%
	\providecommand \enquote  [1]{``#1''}%
	\providecommand \bibnamefont  [1]{#1}%
	\providecommand \bibfnamefont [1]{#1}%
	\providecommand \citenamefont [1]{#1}%
	\providecommand \href@noop [0]{\@secondoftwo}%
	\providecommand \href [0]{\begingroup \@sanitize@url \@href}%
	\providecommand \@href[1]{\@@startlink{#1}\@@href}%
	\providecommand \@@href[1]{\endgroup#1\@@endlink}%
	\providecommand \@sanitize@url [0]{\catcode `\\12\catcode `\$12\catcode
		`\&12\catcode `\#12\catcode `\^12\catcode `\_12\catcode `\%12\relax}%
	\providecommand \@@startlink[1]{}%
	\providecommand \@@endlink[0]{}%
	\providecommand \url  [0]{\begingroup\@sanitize@url \@url }%
	\providecommand \@url [1]{\endgroup\@href {#1}{\urlprefix }}%
	\providecommand \urlprefix  [0]{URL }%
	\providecommand \Eprint [0]{\href }%
	\providecommand \doibase [0]{https://doi.org/}%
	\providecommand \selectlanguage [0]{\@gobble}%
	\providecommand \bibinfo  [0]{\@secondoftwo}%
	\providecommand \bibfield  [0]{\@secondoftwo}%
	\providecommand \translation [1]{[#1]}%
	\providecommand \BibitemOpen [0]{}%
	\providecommand \bibitemStop [0]{}%
	\providecommand \bibitemNoStop [0]{.\EOS\space}%
	\providecommand \EOS [0]{\spacefactor3000\relax}%
	\providecommand \BibitemShut  [1]{\csname bibitem#1\endcsname}%
	\let\auto@bib@innerbib\@empty
	%</preamble>
	\bibitem [{\citenamefont {Feynman}\ \emph {et~al.}(1965)\citenamefont
		{Feynman}, \citenamefont {Leighton},\ and\ \citenamefont {Sands}}]{fey65}%
	\BibitemOpen
	\bibfield  {author} {\bibinfo {author} {\bibfnamefont {R.}~\bibnamefont
			{Feynman}}, \bibinfo {author} {\bibfnamefont {R.}~\bibnamefont {Leighton}},\
		and\ \bibinfo {author} {\bibfnamefont {M.}~\bibnamefont {Sands}},\
	}\href@noop {} {\emph {\bibinfo {title} {The Feynman Lectures on Physics}}}\
	(\bibinfo  {publisher} {Addison-Wesley, Reading},\ \bibinfo {year}
	{1965})\BibitemShut {NoStop}%
	\bibitem [{\citenamefont {Peres}(2002)}]{per02}%
	\BibitemOpen
	\bibfield  {author} {\bibinfo {author} {\bibfnamefont {A.}~\bibnamefont
			{Peres}},\ }\href@noop {} {\emph {\bibinfo {title} {Quantum Theory: Concepts
				and Methods}}}\ (\bibinfo  {publisher} {Kluwer Academic Publishers, New
		York},\ \bibinfo {year} {2002})\BibitemShut {NoStop}%
	\bibitem [{\citenamefont {Sorkin}(1994)}]{Sorkin:1994}%
	\BibitemOpen
	\bibfield  {author} {\bibinfo {author} {\bibfnamefont {R.~D.}\ \bibnamefont
			{Sorkin}},\ }\bibfield  {title} {\bibinfo {title} {Quantum mechanics as
			quantum measure theory},\ }\href {https://doi.org/10.1142/S021773239400294X}
	{\bibfield  {journal} {\bibinfo  {journal} {Mod. Phys. Lett. A}\ }\textbf
		{\bibinfo {volume} {09}},\ \bibinfo {pages} {3119} (\bibinfo {year}
		{1994})}\BibitemShut {NoStop}%
	\bibitem [{\citenamefont {Born}(1926)}]{Born:1926}%
	\BibitemOpen
	\bibfield  {author} {\bibinfo {author} {\bibfnamefont {M.}~\bibnamefont
			{Born}},\ }\bibfield  {title} {\bibinfo {title} {Zur {Q}uantenmechanik der
			{S}to{\ss}vorg{\"a}nge},\ }\href {https://doi.org/10.1007/BF01397477}
	{\bibfield  {journal} {\bibinfo  {journal} {Z. Phys.}\ }\textbf {\bibinfo
			{volume} {37}},\ \bibinfo {pages} {863} (\bibinfo {year} {1926})}\BibitemShut
	{NoStop}%
	\bibitem [{\citenamefont {Niestegge}(2013)}]{Niestegge:2013}%
	\BibitemOpen
	\bibfield  {author} {\bibinfo {author} {\bibfnamefont {G.}~\bibnamefont
			{Niestegge}},\ }\bibfield  {title} {\bibinfo {title} {Three-slit experiments
			and quantum nonlocality},\ }\href {https://doi.org/10.1007/s10701-013-9719-3}
	{\bibfield  {journal} {\bibinfo  {journal} {Found. Phys.}\ }\textbf {\bibinfo
			{volume} {43}},\ \bibinfo {pages} {805} (\bibinfo {year} {2013})}\BibitemShut
	{NoStop}%
	\bibitem [{\citenamefont {Daki{\'{c}}}\ \emph {et~al.}(2014)\citenamefont
		{Daki{\'{c}}}, \citenamefont {Paterek},\ and\ \citenamefont
		{Brukner}}]{Dakic:2014}%
	\BibitemOpen
	\bibfield  {author} {\bibinfo {author} {\bibfnamefont {B.}~\bibnamefont
			{Daki{\'{c}}}}, \bibinfo {author} {\bibfnamefont {T.}~\bibnamefont
			{Paterek}},\ and\ \bibinfo {author} {\bibfnamefont {{\v{C}}.}~\bibnamefont
			{Brukner}},\ }\bibfield  {title} {\bibinfo {title} {Density cubes and
			higher-order interference theories},\ }\href
	{https://doi.org/10.1088/1367-2630/16/2/023028} {\bibfield  {journal}
		{\bibinfo  {journal} {New J. Phys.}\ }\textbf {\bibinfo {volume} {16}},\
		\bibinfo {pages} {023028} (\bibinfo {year} {2014})}\BibitemShut {NoStop}%
	\bibitem [{\citenamefont {Henson}(2015)}]{Henson:2015}%
	\BibitemOpen
	\bibfield  {author} {\bibinfo {author} {\bibfnamefont {J.}~\bibnamefont
			{Henson}},\ }\bibfield  {title} {\bibinfo {title} {Bounding quantum
			contextuality with lack of third-order interference},\ }\href
	{https://doi.org/10.1103/PhysRevLett.114.220403} {\bibfield  {journal}
		{\bibinfo  {journal} {Phys. Rev. Lett.}\ }\textbf {\bibinfo {volume} {114}},\
		\bibinfo {pages} {220403} (\bibinfo {year} {2015})}\BibitemShut {NoStop}%
	\bibitem [{\citenamefont {Zhao}\ \emph {et~al.}(2018)\citenamefont {Zhao},
		\citenamefont {Mondal}, \citenamefont {Markiewicz}, \citenamefont
		{Rutkowski}, \citenamefont {Daki\ifmmode~\acute{c}\else \'{c}\fi{}},
		\citenamefont {Laskowski},\ and\ \citenamefont {Paterek}}]{Zhao:2018}%
	\BibitemOpen
	\bibfield  {author} {\bibinfo {author} {\bibfnamefont {Z.}~\bibnamefont
			{Zhao}}, \bibinfo {author} {\bibfnamefont {S.}~\bibnamefont {Mondal}},
		\bibinfo {author} {\bibfnamefont {M.}~\bibnamefont {Markiewicz}}, \bibinfo
		{author} {\bibfnamefont {A.}~\bibnamefont {Rutkowski}}, \bibinfo {author}
		{\bibfnamefont {B.}~\bibnamefont {Daki\ifmmode~\acute{c}\else \'{c}\fi{}}},
		\bibinfo {author} {\bibfnamefont {W.}~\bibnamefont {Laskowski}},\ and\
		\bibinfo {author} {\bibfnamefont {T.}~\bibnamefont {Paterek}},\ }\bibfield
	{title} {\bibinfo {title} {Paradoxical consequences of multipath coherence:
			Perfect interaction-free measurements},\ }\href
	{https://doi.org/10.1103/PhysRevA.98.022108} {\bibfield  {journal} {\bibinfo
			{journal} {Phys. Rev. A}\ }\textbf {\bibinfo {volume} {98}},\ \bibinfo
		{pages} {022108} (\bibinfo {year} {2018})}\BibitemShut {NoStop}%
	\bibitem [{\citenamefont {Sinha}\ \emph {et~al.}(2010)\citenamefont {Sinha},
		\citenamefont {Couteau}, \citenamefont {Jennewein}, \citenamefont
		{Laflamme},\ and\ \citenamefont {Weihs}}]{Sinha:2010}%
	\BibitemOpen
	\bibfield  {author} {\bibinfo {author} {\bibfnamefont {U.}~\bibnamefont
			{Sinha}}, \bibinfo {author} {\bibfnamefont {C.}~\bibnamefont {Couteau}},
		\bibinfo {author} {\bibfnamefont {T.}~\bibnamefont {Jennewein}}, \bibinfo
		{author} {\bibfnamefont {R.}~\bibnamefont {Laflamme}},\ and\ \bibinfo
		{author} {\bibfnamefont {G.}~\bibnamefont {Weihs}},\ }\bibfield  {title}
	{\bibinfo {title} {Ruling out multi-order interference in quantum
			mechanics},\ }\href {https://doi.org/10.1126/science.1190545} {\bibfield
		{journal} {\bibinfo  {journal} {Science}\ }\textbf {\bibinfo {volume}
			{329}},\ \bibinfo {pages} {418} (\bibinfo {year} {2010})}\BibitemShut
	{NoStop}%
	\bibitem [{\citenamefont {Hickmann}\ \emph {et~al.}(2011)\citenamefont
		{Hickmann}, \citenamefont {Fonseca},\ and\ \citenamefont
		{Jesus-Silva}}]{Hickmann:2011}%
	\BibitemOpen
	\bibfield  {author} {\bibinfo {author} {\bibfnamefont {J.~M.}\ \bibnamefont
			{Hickmann}}, \bibinfo {author} {\bibfnamefont {E.~J.~S.}\ \bibnamefont
			{Fonseca}},\ and\ \bibinfo {author} {\bibfnamefont {A.~J.}\ \bibnamefont
			{Jesus-Silva}},\ }\bibfield  {title} {\bibinfo {title} {Born's rule and the
			interference of photons with orbital angular momentum by a triangular slit},\
	}\href {https://doi.org/10.1209/0295-5075/96/64006} {\bibfield  {journal}
		{\bibinfo  {journal} {{EPL}}\ }\textbf {\bibinfo {volume} {96}},\ \bibinfo
		{pages} {64006} (\bibinfo {year} {2011})}\BibitemShut {NoStop}%
	\bibitem [{\citenamefont {S{\"o}llner}\ \emph {et~al.}(2012)\citenamefont
		{S{\"o}llner}, \citenamefont {Gsch{\"o}sser}, \citenamefont {Mai},
		\citenamefont {Pressl}, \citenamefont {V{\"o}r{\"o}s},\ and\ \citenamefont
		{Weihs}}]{Sollner:2012}%
	\BibitemOpen
	\bibfield  {author} {\bibinfo {author} {\bibfnamefont {I.}~\bibnamefont
			{S{\"o}llner}}, \bibinfo {author} {\bibfnamefont {B.}~\bibnamefont
			{Gsch{\"o}sser}}, \bibinfo {author} {\bibfnamefont {P.}~\bibnamefont {Mai}},
		\bibinfo {author} {\bibfnamefont {B.}~\bibnamefont {Pressl}}, \bibinfo
		{author} {\bibfnamefont {Z.}~\bibnamefont {V{\"o}r{\"o}s}},\ and\ \bibinfo
		{author} {\bibfnamefont {G.}~\bibnamefont {Weihs}},\ }\bibfield  {title}
	{\bibinfo {title} {Testing {B}orn's rule in quantum mechanics for three
			mutually exclusive events},\ }\href
	{https://doi.org/10.1007/s10701-011-9597-5} {\bibfield  {journal} {\bibinfo
			{journal} {Found. Phys.}\ }\textbf {\bibinfo {volume} {42}},\ \bibinfo
		{pages} {742} (\bibinfo {year} {2012})}\BibitemShut {NoStop}%
	\bibitem [{\citenamefont {Kauten}\ \emph {et~al.}(2017)\citenamefont {Kauten},
		\citenamefont {Keil}, \citenamefont {Kaufmann}, \citenamefont {Pressl},
		\citenamefont {Brukner},\ and\ \citenamefont {Weihs}}]{Kauten:2017}%
	\BibitemOpen
	\bibfield  {author} {\bibinfo {author} {\bibfnamefont {T.}~\bibnamefont
			{Kauten}}, \bibinfo {author} {\bibfnamefont {R.}~\bibnamefont {Keil}},
		\bibinfo {author} {\bibfnamefont {T.}~\bibnamefont {Kaufmann}}, \bibinfo
		{author} {\bibfnamefont {B.}~\bibnamefont {Pressl}}, \bibinfo {author}
		{\bibfnamefont {{\v C}.}~\bibnamefont {Brukner}},\ and\ \bibinfo {author}
		{\bibfnamefont {G.}~\bibnamefont {Weihs}},\ }\bibfield  {title} {\bibinfo
		{title} {Obtaining tight bounds on higher-order interferences with a 5-path
			interferometer},\ }\href {http://stacks.iop.org/1367-2630/19/i=3/a=033017}
	{\bibfield  {journal} {\bibinfo  {journal} {New J. Phys.}\ }\textbf {\bibinfo
			{volume} {19}},\ \bibinfo {pages} {033017} (\bibinfo {year}
		{2017})}\BibitemShut {NoStop}%
	\bibitem [{\citenamefont {Rengaraj}\ \emph {et~al.}(2018)\citenamefont
		{Rengaraj}, \citenamefont {Prathwiraj}, \citenamefont {Sahoo}, \citenamefont
		{Somashekhar},\ and\ \citenamefont {Sinha}}]{Rengaraj:2018}%
	\BibitemOpen
	\bibfield  {author} {\bibinfo {author} {\bibfnamefont {G.}~\bibnamefont
			{Rengaraj}}, \bibinfo {author} {\bibfnamefont {U.}~\bibnamefont
			{Prathwiraj}}, \bibinfo {author} {\bibfnamefont {S.~N.}\ \bibnamefont
			{Sahoo}}, \bibinfo {author} {\bibfnamefont {R.}~\bibnamefont {Somashekhar}},\
		and\ \bibinfo {author} {\bibfnamefont {U.}~\bibnamefont {Sinha}},\ }\bibfield
	{title} {\bibinfo {title} {Measuring the deviation from the superposition
			principle in interference experiments},\ }\href
	{https://doi.org/10.1088/1367-2630/aac92c} {\bibfield  {journal} {\bibinfo
			{journal} {New Journal of Physics}\ }\textbf {\bibinfo {volume} {20}},\
		\bibinfo {pages} {063049} (\bibinfo {year} {2018})}\BibitemShut {NoStop}%
	\bibitem [{\citenamefont {Cotter}\ \emph {et~al.}(2017)\citenamefont {Cotter},
		\citenamefont {Brand}, \citenamefont {Knobloch}, \citenamefont {Lilach},
		\citenamefont {Cheshnovsky},\ and\ \citenamefont {Arndt}}]{Cotter:2017}%
	\BibitemOpen
	\bibfield  {author} {\bibinfo {author} {\bibfnamefont {J.~P.}\ \bibnamefont
			{Cotter}}, \bibinfo {author} {\bibfnamefont {C.}~\bibnamefont {Brand}},
		\bibinfo {author} {\bibfnamefont {C.}~\bibnamefont {Knobloch}}, \bibinfo
		{author} {\bibfnamefont {Y.}~\bibnamefont {Lilach}}, \bibinfo {author}
		{\bibfnamefont {O.}~\bibnamefont {Cheshnovsky}},\ and\ \bibinfo {author}
		{\bibfnamefont {M.}~\bibnamefont {Arndt}},\ }\bibfield  {title} {\bibinfo
		{title} {In search of multipath interference using large molecules},\ }\href
	{https://advances.sciencemag.org/content/3/8/e1602478} {\bibfield  {journal}
		{\bibinfo  {journal} {Sci. Adv.}\ }\textbf {\bibinfo {volume} {3}},\ \bibinfo
		{pages} {e1602478} (\bibinfo {year} {2017})}\BibitemShut {NoStop}%
	\bibitem [{\citenamefont {Barnea}\ \emph {et~al.}(2018)\citenamefont {Barnea},
		\citenamefont {Cheshnovsky},\ and\ \citenamefont {Even}}]{Barnea:2018}%
	\BibitemOpen
	\bibfield  {author} {\bibinfo {author} {\bibfnamefont {A.~R.}\ \bibnamefont
			{Barnea}}, \bibinfo {author} {\bibfnamefont {O.}~\bibnamefont
			{Cheshnovsky}},\ and\ \bibinfo {author} {\bibfnamefont {U.}~\bibnamefont
			{Even}},\ }\bibfield  {title} {\bibinfo {title} {Matter-wave diffraction
			approaching limits predicted by feynman path integrals for multipath
			interference},\ }\href {https://doi.org/10.1103/PhysRevA.97.023601}
	{\bibfield  {journal} {\bibinfo  {journal} {Phys. Rev. A}\ }\textbf {\bibinfo
			{volume} {97}},\ \bibinfo {pages} {023601} (\bibinfo {year}
		{2018})}\BibitemShut {NoStop}%
	\bibitem [{\citenamefont {Park}\ \emph {et~al.}(2012)\citenamefont {Park},
		\citenamefont {Moussa},\ and\ \citenamefont {Laflamme}}]{Park:2012}%
	\BibitemOpen
	\bibfield  {author} {\bibinfo {author} {\bibfnamefont {D.~K.}\ \bibnamefont
			{Park}}, \bibinfo {author} {\bibfnamefont {O.}~\bibnamefont {Moussa}},\ and\
		\bibinfo {author} {\bibfnamefont {R.}~\bibnamefont {Laflamme}},\ }\bibfield
	{title} {\bibinfo {title} {Three path interference using nuclear magnetic
			resonance: a test of the consistency of {B}orn's rule},\ }\href
	{https://doi.org/10.1088/1367-2630/14/11/113025} {\bibfield  {journal}
		{\bibinfo  {journal} {New J. Phys.}\ }\textbf {\bibinfo {volume} {14}},\
		\bibinfo {pages} {113025} (\bibinfo {year} {2012})}\BibitemShut {NoStop}%
	\bibitem [{\citenamefont {Jin}\ \emph {et~al.}(2017)\citenamefont {Jin},
		\citenamefont {Liu}, \citenamefont {Geng}, \citenamefont {Huang},
		\citenamefont {Ma}, \citenamefont {Shi}, \citenamefont {Duan}, \citenamefont
		{Shi}, \citenamefont {Rong},\ and\ \citenamefont {Du}}]{Jin:2017}%
	\BibitemOpen
	\bibfield  {author} {\bibinfo {author} {\bibfnamefont {F.}~\bibnamefont
			{Jin}}, \bibinfo {author} {\bibfnamefont {Y.}~\bibnamefont {Liu}}, \bibinfo
		{author} {\bibfnamefont {J.}~\bibnamefont {Geng}}, \bibinfo {author}
		{\bibfnamefont {P.}~\bibnamefont {Huang}}, \bibinfo {author} {\bibfnamefont
			{W.}~\bibnamefont {Ma}}, \bibinfo {author} {\bibfnamefont {M.}~\bibnamefont
			{Shi}}, \bibinfo {author} {\bibfnamefont {C.-K.}\ \bibnamefont {Duan}},
		\bibinfo {author} {\bibfnamefont {F.}~\bibnamefont {Shi}}, \bibinfo {author}
		{\bibfnamefont {X.}~\bibnamefont {Rong}},\ and\ \bibinfo {author}
		{\bibfnamefont {J.}~\bibnamefont {Du}},\ }\bibfield  {title} {\bibinfo
		{title} {Experimental test of {B}orn's rule by inspecting third-order quantum
			interference on a single spin in solids},\ }\href
	{https://doi.org/10.1103/PhysRevA.95.012107} {\bibfield  {journal} {\bibinfo
			{journal} {Phys. Rev. A}\ }\textbf {\bibinfo {volume} {95}},\ \bibinfo
		{pages} {012107} (\bibinfo {year} {2017})}\BibitemShut {NoStop}%
	\bibitem [{\citenamefont {Shull}\ \emph {et~al.}(1980)\citenamefont {Shull},
		\citenamefont {Atwood}, \citenamefont {Arthur},\ and\ \citenamefont
		{Horne}}]{Shull:1980}%
	\BibitemOpen
	\bibfield  {author} {\bibinfo {author} {\bibfnamefont {C.~G.}\ \bibnamefont
			{Shull}}, \bibinfo {author} {\bibfnamefont {D.~K.}\ \bibnamefont {Atwood}},
		\bibinfo {author} {\bibfnamefont {J.}~\bibnamefont {Arthur}},\ and\ \bibinfo
		{author} {\bibfnamefont {M.~A.}\ \bibnamefont {Horne}},\ }\bibfield  {title}
	{\bibinfo {title} {Search for a nonlinear variant of the {S}chr\"odinger
			equation by neutron interferometry},\ }\href
	{https://doi.org/10.1103/PhysRevLett.44.765} {\bibfield  {journal} {\bibinfo
			{journal} {Phys. Rev. Lett.}\ }\textbf {\bibinfo {volume} {44}},\ \bibinfo
		{pages} {765} (\bibinfo {year} {1980})}\BibitemShut {NoStop}%
	\bibitem [{\citenamefont {G\"ahler}\ \emph {et~al.}(1981)\citenamefont
		{G\"ahler}, \citenamefont {Klein},\ and\ \citenamefont
		{Zeilinger}}]{Gahler:1981}%
	\BibitemOpen
	\bibfield  {author} {\bibinfo {author} {\bibfnamefont {R.}~\bibnamefont
			{G\"ahler}}, \bibinfo {author} {\bibfnamefont {A.~G.}\ \bibnamefont
			{Klein}},\ and\ \bibinfo {author} {\bibfnamefont {A.}~\bibnamefont
			{Zeilinger}},\ }\bibfield  {title} {\bibinfo {title} {Neutron optical tests
			of nonlinear wave mechanics},\ }\href
	{https://doi.org/10.1103/PhysRevA.23.1611} {\bibfield  {journal} {\bibinfo
			{journal} {Phys. Rev. A}\ }\textbf {\bibinfo {volume} {23}},\ \bibinfo
		{pages} {1611} (\bibinfo {year} {1981})}\BibitemShut {NoStop}%
	\bibitem [{\citenamefont {Bollinger}\ \emph {et~al.}(1989)\citenamefont
		{Bollinger}, \citenamefont {Heinzen}, \citenamefont {Itano}, \citenamefont
		{Gilbert},\ and\ \citenamefont {Wineland}}]{Bollinger:1989}%
	\BibitemOpen
	\bibfield  {author} {\bibinfo {author} {\bibfnamefont {J.~J.}\ \bibnamefont
			{Bollinger}}, \bibinfo {author} {\bibfnamefont {D.~J.}\ \bibnamefont
			{Heinzen}}, \bibinfo {author} {\bibfnamefont {W.~M.}\ \bibnamefont {Itano}},
		\bibinfo {author} {\bibfnamefont {S.~L.}\ \bibnamefont {Gilbert}},\ and\
		\bibinfo {author} {\bibfnamefont {D.~J.}\ \bibnamefont {Wineland}},\
	}\bibfield  {title} {\bibinfo {title} {Test of the linearity of quantum
			mechanics by rf spectroscopy of the $^{9}\mathrm{Be}^{+}$ ground state},\
	}\href {https://link.aps.org/doi/10.1103/PhysRevLett.63.1031} {\bibfield
		{journal} {\bibinfo  {journal} {Phys. Rev. Lett.}\ }\textbf {\bibinfo
			{volume} {63}},\ \bibinfo {pages} {1031} (\bibinfo {year}
		{1989})}\BibitemShut {NoStop}%
	\bibitem [{\citenamefont {Chupp}\ and\ \citenamefont
		{Hoare}(1990)}]{Chupp:1990}%
	\BibitemOpen
	\bibfield  {author} {\bibinfo {author} {\bibfnamefont {T.~E.}\ \bibnamefont
			{Chupp}}\ and\ \bibinfo {author} {\bibfnamefont {R.~J.}\ \bibnamefont
			{Hoare}},\ }\bibfield  {title} {\bibinfo {title} {Coherence in freely
			precessing $^{21}\mathrm{Ne}$ and a test of linearity of quantum mechanics},\
	}\href {https://link.aps.org/doi/10.1103/PhysRevLett.64.2261} {\bibfield
		{journal} {\bibinfo  {journal} {Phys. Rev. Lett.}\ }\textbf {\bibinfo
			{volume} {64}},\ \bibinfo {pages} {2261} (\bibinfo {year}
		{1990})}\BibitemShut {NoStop}%
	\bibitem [{\citenamefont {Walsworth}\ \emph {et~al.}(1990)\citenamefont
		{Walsworth}, \citenamefont {Silvera}, \citenamefont {Mattison},\ and\
		\citenamefont {Vessot}}]{Walsworth:1990}%
	\BibitemOpen
	\bibfield  {author} {\bibinfo {author} {\bibfnamefont {R.~L.}\ \bibnamefont
			{Walsworth}}, \bibinfo {author} {\bibfnamefont {I.~F.}\ \bibnamefont
			{Silvera}}, \bibinfo {author} {\bibfnamefont {E.~M.}\ \bibnamefont
			{Mattison}},\ and\ \bibinfo {author} {\bibfnamefont {R.~F.~C.}\ \bibnamefont
			{Vessot}},\ }\bibfield  {title} {\bibinfo {title} {Test of the linearity of
			quantum mechanics in an atomic system with a hydrogen maser},\ }\href
	{https://doi.org/10.1103/PhysRevLett.64.2599} {\bibfield  {journal} {\bibinfo
			{journal} {Phys. Rev. Lett.}\ }\textbf {\bibinfo {volume} {64}},\ \bibinfo
		{pages} {2599} (\bibinfo {year} {1990})}\BibitemShut {NoStop}%
	\bibitem [{\citenamefont {Majumder}\ \emph {et~al.}(1990)\citenamefont
		{Majumder}, \citenamefont {Venema}, \citenamefont {Lamoreaux}, \citenamefont
		{Heckel},\ and\ \citenamefont {Fortson}}]{Majumder:1990}%
	\BibitemOpen
	\bibfield  {author} {\bibinfo {author} {\bibfnamefont {P.~K.}\ \bibnamefont
			{Majumder}}, \bibinfo {author} {\bibfnamefont {B.~J.}\ \bibnamefont
			{Venema}}, \bibinfo {author} {\bibfnamefont {S.~K.}\ \bibnamefont
			{Lamoreaux}}, \bibinfo {author} {\bibfnamefont {B.~R.}\ \bibnamefont
			{Heckel}},\ and\ \bibinfo {author} {\bibfnamefont {E.~N.}\ \bibnamefont
			{Fortson}},\ }\bibfield  {title} {\bibinfo {title} {Test of the linearity of
			quantum mechanics in optically pumped $^{201}\mathrm{Hg}$},\ }\href
	{https://doi.org/10.1103/PhysRevLett.65.2931} {\bibfield  {journal} {\bibinfo
			{journal} {Phys. Rev. Lett.}\ }\textbf {\bibinfo {volume} {65}},\ \bibinfo
		{pages} {2931} (\bibinfo {year} {1990})}\BibitemShut {NoStop}%
	\bibitem [{\citenamefont {Vinante}\ \emph {et~al.}(2017)\citenamefont
		{Vinante}, \citenamefont {Mezzena}, \citenamefont {Falferi}, \citenamefont
		{Carlesso},\ and\ \citenamefont {Bassi}}]{Vinante:2017}%
	\BibitemOpen
	\bibfield  {author} {\bibinfo {author} {\bibfnamefont {A.}~\bibnamefont
			{Vinante}}, \bibinfo {author} {\bibfnamefont {R.}~\bibnamefont {Mezzena}},
		\bibinfo {author} {\bibfnamefont {P.}~\bibnamefont {Falferi}}, \bibinfo
		{author} {\bibfnamefont {M.}~\bibnamefont {Carlesso}},\ and\ \bibinfo
		{author} {\bibfnamefont {A.}~\bibnamefont {Bassi}},\ }\bibfield  {title}
	{\bibinfo {title} {Improved noninterferometric test of collapse models using
			ultracold cantilevers},\ }\href
	{https://link.aps.org/doi/10.1103/PhysRevLett.119.110401} {\bibfield
		{journal} {\bibinfo  {journal} {Phys. Rev. Lett.}\ }\textbf {\bibinfo
			{volume} {119}},\ \bibinfo {pages} {110401} (\bibinfo {year}
		{2017})}\BibitemShut {NoStop}%
	\bibitem [{\citenamefont {Hong}\ \emph {et~al.}(1987)\citenamefont {Hong},
		\citenamefont {Ou},\ and\ \citenamefont {Mandel}}]{Hong:1987}%
	\BibitemOpen
	\bibfield  {author} {\bibinfo {author} {\bibfnamefont {C.~K.}\ \bibnamefont
			{Hong}}, \bibinfo {author} {\bibfnamefont {Z.~Y.}\ \bibnamefont {Ou}},\ and\
		\bibinfo {author} {\bibfnamefont {L.}~\bibnamefont {Mandel}},\ }\bibfield
	{title} {\bibinfo {title} {Measurement of subpicosecond time intervals
			between two photons by interference},\ }\href
	{https://doi.org/10.1103/PhysRevLett.59.2044} {\bibfield  {journal} {\bibinfo
			{journal} {Phys. Rev. Lett.}\ }\textbf {\bibinfo {volume} {59}},\ \bibinfo
		{pages} {2044} (\bibinfo {year} {1987})}\BibitemShut {NoStop}%
	\bibitem [{\citenamefont {Glauber}(1963)}]{Glauber:1963}%
	\BibitemOpen
	\bibfield  {author} {\bibinfo {author} {\bibfnamefont {R.~J.}\ \bibnamefont
			{Glauber}},\ }\bibfield  {title} {\bibinfo {title} {The quantum theory of
			optical coherence},\ }\href {https://doi.org/10.1103/PhysRev.130.2529}
	{\bibfield  {journal} {\bibinfo  {journal} {Phys. Rev.}\ }\textbf {\bibinfo
			{volume} {130}},\ \bibinfo {pages} {2529} (\bibinfo {year}
		{1963})}\BibitemShut {NoStop}%
	\bibitem [{\citenamefont {Pan}\ \emph {et~al.}(2012)\citenamefont {Pan},
		\citenamefont {Chen}, \citenamefont {Lu}, \citenamefont {Weinfurter},
		\citenamefont {Zeilinger},\ and\ \citenamefont {\ifmmode~\dot{Z}\else
			\.{Z}\fi{}ukowski}}]{Pan:2012}%
	\BibitemOpen
	\bibfield  {author} {\bibinfo {author} {\bibfnamefont {J.-W.}\ \bibnamefont
			{Pan}}, \bibinfo {author} {\bibfnamefont {Z.-B.}\ \bibnamefont {Chen}},
		\bibinfo {author} {\bibfnamefont {C.-Y.}\ \bibnamefont {Lu}}, \bibinfo
		{author} {\bibfnamefont {H.}~\bibnamefont {Weinfurter}}, \bibinfo {author}
		{\bibfnamefont {A.}~\bibnamefont {Zeilinger}},\ and\ \bibinfo {author}
		{\bibfnamefont {M.}~\bibnamefont {\ifmmode~\dot{Z}\else \.{Z}\fi{}ukowski}},\
	}\bibfield  {title} {\bibinfo {title} {Multiphoton entanglement and
			interferometry},\ }\href {https://doi.org/10.1103/RevModPhys.84.777}
	{\bibfield  {journal} {\bibinfo  {journal} {Rev. Mod. Phys.}\ }\textbf
		{\bibinfo {volume} {84}},\ \bibinfo {pages} {777} (\bibinfo {year}
		{2012})}\BibitemShut {NoStop}%
	\bibitem [{\citenamefont {Tichy}(2014)}]{Tichy:2014}%
	\BibitemOpen
	\bibfield  {author} {\bibinfo {author} {\bibfnamefont {M.~C.}\ \bibnamefont
			{Tichy}},\ }\bibfield  {title} {\bibinfo {title} {Interference of identical
			particles from entanglement to boson-sampling},\ }\href
	{http://stacks.iop.org/0953-4075/47/i=10/a=103001} {\bibfield  {journal}
		{\bibinfo  {journal} {J. Phys. B At. Mol. Opt. Phys.}\ }\textbf {\bibinfo
			{volume} {47}},\ \bibinfo {pages} {103001} (\bibinfo {year}
		{2014})}\BibitemShut {NoStop}%
	\bibitem [{\citenamefont {Agne}\ \emph {et~al.}(2017)\citenamefont {Agne},
		\citenamefont {Kauten}, \citenamefont {Jin}, \citenamefont {Meyer-Scott},
		\citenamefont {Salvail}, \citenamefont {Hamel}, \citenamefont {Resch},
		\citenamefont {Weihs},\ and\ \citenamefont {Jennewein}}]{Agne:2017}%
	\BibitemOpen
	\bibfield  {author} {\bibinfo {author} {\bibfnamefont {S.}~\bibnamefont
			{Agne}}, \bibinfo {author} {\bibfnamefont {T.}~\bibnamefont {Kauten}},
		\bibinfo {author} {\bibfnamefont {J.}~\bibnamefont {Jin}}, \bibinfo {author}
		{\bibfnamefont {E.}~\bibnamefont {Meyer-Scott}}, \bibinfo {author}
		{\bibfnamefont {J.~Z.}\ \bibnamefont {Salvail}}, \bibinfo {author}
		{\bibfnamefont {D.~R.}\ \bibnamefont {Hamel}}, \bibinfo {author}
		{\bibfnamefont {K.~J.}\ \bibnamefont {Resch}}, \bibinfo {author}
		{\bibfnamefont {G.}~\bibnamefont {Weihs}},\ and\ \bibinfo {author}
		{\bibfnamefont {T.}~\bibnamefont {Jennewein}},\ }\bibfield  {title} {\bibinfo
		{title} {Observation of genuine three-photon interference},\ }\href
	{https://doi.org/10.1103/PhysRevLett.118.153602} {\bibfield  {journal}
		{\bibinfo  {journal} {Phys. Rev. Lett.}\ }\textbf {\bibinfo {volume} {118}},\
		\bibinfo {pages} {153602} (\bibinfo {year} {2017})}\BibitemShut {NoStop}%
	\bibitem [{\citenamefont {Menssen}\ \emph {et~al.}(2017)\citenamefont
		{Menssen}, \citenamefont {Jones}, \citenamefont {Metcalf}, \citenamefont
		{Tichy}, \citenamefont {Barz}, \citenamefont {Kolthammer},\ and\
		\citenamefont {Walmsley}}]{Menssen:2017}%
	\BibitemOpen
	\bibfield  {author} {\bibinfo {author} {\bibfnamefont {A.~J.}\ \bibnamefont
			{Menssen}}, \bibinfo {author} {\bibfnamefont {A.~E.}\ \bibnamefont {Jones}},
		\bibinfo {author} {\bibfnamefont {B.~J.}\ \bibnamefont {Metcalf}}, \bibinfo
		{author} {\bibfnamefont {M.~C.}\ \bibnamefont {Tichy}}, \bibinfo {author}
		{\bibfnamefont {S.}~\bibnamefont {Barz}}, \bibinfo {author} {\bibfnamefont
			{W.~S.}\ \bibnamefont {Kolthammer}},\ and\ \bibinfo {author} {\bibfnamefont
			{I.~A.}\ \bibnamefont {Walmsley}},\ }\bibfield  {title} {\bibinfo {title}
		{Distinguishability and many-particle interference},\ }\href
	{https://doi.org/10.1103/PhysRevLett.118.153603} {\bibfield  {journal}
		{\bibinfo  {journal} {Phys. Rev. Lett.}\ }\textbf {\bibinfo {volume} {118}},\
		\bibinfo {pages} {153603} (\bibinfo {year} {2017})}\BibitemShut {NoStop}%
	\bibitem [{\citenamefont {Giovannetti}\ \emph {et~al.}(2011)\citenamefont
		{Giovannetti}, \citenamefont {Lloyd},\ and\ \citenamefont
		{Maccone}}]{Giovannetti:2011}%
	\BibitemOpen
	\bibfield  {author} {\bibinfo {author} {\bibfnamefont {V.}~\bibnamefont
			{Giovannetti}}, \bibinfo {author} {\bibfnamefont {S.}~\bibnamefont {Lloyd}},\
		and\ \bibinfo {author} {\bibfnamefont {L.}~\bibnamefont {Maccone}},\
	}\bibfield  {title} {\bibinfo {title} {Advances in quantum metrology},\
	}\href {https://doi.org/10.1038/nphoton.2011.35} {\bibfield  {journal}
		{\bibinfo  {journal} {Nat. Photonics}\ }\textbf {\bibinfo {volume} {5}},\
		\bibinfo {pages} {222} (\bibinfo {year} {2011})}\BibitemShut {NoStop}%
	\bibitem [{\citenamefont {Su}\ \emph {et~al.}(2017)\citenamefont {Su},
		\citenamefont {Li}, \citenamefont {Rohde}, \citenamefont {Huang},
		\citenamefont {Wang}, \citenamefont {Li}, \citenamefont {Liu}, \citenamefont
		{Dowling}, \citenamefont {Lu},\ and\ \citenamefont {Pan}}]{Su:2017}%
	\BibitemOpen
	\bibfield  {author} {\bibinfo {author} {\bibfnamefont {Z.-E.}\ \bibnamefont
			{Su}}, \bibinfo {author} {\bibfnamefont {Y.}~\bibnamefont {Li}}, \bibinfo
		{author} {\bibfnamefont {P.~P.}\ \bibnamefont {Rohde}}, \bibinfo {author}
		{\bibfnamefont {H.-L.}\ \bibnamefont {Huang}}, \bibinfo {author}
		{\bibfnamefont {X.-L.}\ \bibnamefont {Wang}}, \bibinfo {author}
		{\bibfnamefont {L.}~\bibnamefont {Li}}, \bibinfo {author} {\bibfnamefont
			{N.-L.}\ \bibnamefont {Liu}}, \bibinfo {author} {\bibfnamefont {J.~P.}\
			\bibnamefont {Dowling}}, \bibinfo {author} {\bibfnamefont {C.-Y.}\
			\bibnamefont {Lu}},\ and\ \bibinfo {author} {\bibfnamefont {J.-W.}\
			\bibnamefont {Pan}},\ }\bibfield  {title} {\bibinfo {title} {Multiphoton
			interference in quantum {F}ourier transform circuits and applications to
			quantum metrology},\ }\href {https://doi.org/10.1103/PhysRevLett.119.080502}
	{\bibfield  {journal} {\bibinfo  {journal} {Phys. Rev. Lett.}\ }\textbf
		{\bibinfo {volume} {119}},\ \bibinfo {pages} {080502} (\bibinfo {year}
		{2017})}\BibitemShut {NoStop}%
	\bibitem [{\citenamefont {Hanbury~Brown}\ and\ \citenamefont
		{Twiss}(1956)}]{Hanbury-Brown:1956}%
	\BibitemOpen
	\bibfield  {author} {\bibinfo {author} {\bibfnamefont {R.}~\bibnamefont
			{Hanbury~Brown}}\ and\ \bibinfo {author} {\bibfnamefont {R.~Q.}\ \bibnamefont
			{Twiss}},\ }\bibfield  {title} {\bibinfo {title} {A test of a new type of
			stellar interferometer on {S}irius},\ }\href
	{https://doi.org/10.1038/1781046a0} {\bibfield  {journal} {\bibinfo
			{journal} {Nature}\ }\textbf {\bibinfo {volume} {178}},\ \bibinfo {pages}
		{1046} (\bibinfo {year} {1956})}\BibitemShut {NoStop}%
	\bibitem [{\citenamefont {Thiel}\ \emph {et~al.}(2007)\citenamefont {Thiel},
		\citenamefont {Bastin}, \citenamefont {Martin}, \citenamefont {Solano},
		\citenamefont {von Zanthier},\ and\ \citenamefont {Agarwal}}]{Thiel:2007}%
	\BibitemOpen
	\bibfield  {author} {\bibinfo {author} {\bibfnamefont {C.}~\bibnamefont
			{Thiel}}, \bibinfo {author} {\bibfnamefont {T.}~\bibnamefont {Bastin}},
		\bibinfo {author} {\bibfnamefont {J.}~\bibnamefont {Martin}}, \bibinfo
		{author} {\bibfnamefont {E.}~\bibnamefont {Solano}}, \bibinfo {author}
		{\bibfnamefont {J.}~\bibnamefont {von Zanthier}},\ and\ \bibinfo {author}
		{\bibfnamefont {G.~S.}\ \bibnamefont {Agarwal}},\ }\bibfield  {title}
	{\bibinfo {title} {Quantum imaging with incoherent photons},\ }\href
	{https://doi.org/10.1103/PhysRevLett.99.133603} {\bibfield  {journal}
		{\bibinfo  {journal} {Phys. Rev. Lett.}\ }\textbf {\bibinfo {volume} {99}},\
		\bibinfo {pages} {133603} (\bibinfo {year} {2007})}\BibitemShut {NoStop}%
	\bibitem [{\citenamefont {Schneider}\ \emph {et~al.}(2018)\citenamefont
		{Schneider}, \citenamefont {Mehringer}, \citenamefont {Mercurio},
		\citenamefont {Wenthaus}, \citenamefont {Classen}, \citenamefont {Brenner},
		\citenamefont {Gorobtsov}, \citenamefont {Benz}, \citenamefont {Bhatti},
		\citenamefont {Bocklage}, \citenamefont {Fischer}, \citenamefont {Lazarev},
		\citenamefont {Obukhov}, \citenamefont {Schlage}, \citenamefont {Skopintsev},
		\citenamefont {Wagner}, \citenamefont {Waldmann}, \citenamefont {Willing},
		\citenamefont {Zaluzhnyy}, \citenamefont {Wurth}, \citenamefont
		{Vartanyants}, \citenamefont {R{\"o}hlsberger},\ and\ \citenamefont {von
			Zanthier}}]{Schneider:2018a}%
	\BibitemOpen
	\bibfield  {author} {\bibinfo {author} {\bibfnamefont {R.}~\bibnamefont
			{Schneider}}, \bibinfo {author} {\bibfnamefont {T.}~\bibnamefont
			{Mehringer}}, \bibinfo {author} {\bibfnamefont {G.}~\bibnamefont {Mercurio}},
		\bibinfo {author} {\bibfnamefont {L.}~\bibnamefont {Wenthaus}}, \bibinfo
		{author} {\bibfnamefont {A.}~\bibnamefont {Classen}}, \bibinfo {author}
		{\bibfnamefont {G.}~\bibnamefont {Brenner}}, \bibinfo {author} {\bibfnamefont
			{O.}~\bibnamefont {Gorobtsov}}, \bibinfo {author} {\bibfnamefont
			{A.}~\bibnamefont {Benz}}, \bibinfo {author} {\bibfnamefont {D.}~\bibnamefont
			{Bhatti}}, \bibinfo {author} {\bibfnamefont {L.}~\bibnamefont {Bocklage}},
		\bibinfo {author} {\bibfnamefont {B.}~\bibnamefont {Fischer}}, \bibinfo
		{author} {\bibfnamefont {S.}~\bibnamefont {Lazarev}}, \bibinfo {author}
		{\bibfnamefont {Y.}~\bibnamefont {Obukhov}}, \bibinfo {author} {\bibfnamefont
			{K.}~\bibnamefont {Schlage}}, \bibinfo {author} {\bibfnamefont
			{P.}~\bibnamefont {Skopintsev}}, \bibinfo {author} {\bibfnamefont
			{J.}~\bibnamefont {Wagner}}, \bibinfo {author} {\bibfnamefont
			{F.}~\bibnamefont {Waldmann}}, \bibinfo {author} {\bibfnamefont
			{S.}~\bibnamefont {Willing}}, \bibinfo {author} {\bibfnamefont
			{I.}~\bibnamefont {Zaluzhnyy}}, \bibinfo {author} {\bibfnamefont
			{W.}~\bibnamefont {Wurth}}, \bibinfo {author} {\bibfnamefont {I.~A.}\
			\bibnamefont {Vartanyants}}, \bibinfo {author} {\bibfnamefont
			{R.}~\bibnamefont {R{\"o}hlsberger}},\ and\ \bibinfo {author} {\bibfnamefont
			{J.}~\bibnamefont {von Zanthier}},\ }\bibfield  {title} {\bibinfo {title}
		{Quantum imaging with incoherently scattered light from a free-electron
			laser},\ }\href {https://doi.org/10.1038/nphys4301} {\bibfield  {journal}
		{\bibinfo  {journal} {Nat. Phys.}\ }\textbf {\bibinfo {volume} {14}},\
		\bibinfo {pages} {126} (\bibinfo {year} {2018})}\BibitemShut {NoStop}%
	\bibitem [{\citenamefont {Aaronson}\ and\ \citenamefont
		{Arkhipov}(2011)}]{Aaronson:2011}%
	\BibitemOpen
	\bibfield  {author} {\bibinfo {author} {\bibfnamefont {S.}~\bibnamefont
			{Aaronson}}\ and\ \bibinfo {author} {\bibfnamefont {A.}~\bibnamefont
			{Arkhipov}},\ }\bibfield  {title} {\bibinfo {title} {The computational
			complexity of linear optics},\ }in\ \href
	{https://doi.org/10.1145/1993636.1993682} {\emph {\bibinfo {booktitle}
			{Proceedings of the 43rd Annual ACM Symposium on Theory of Computing}}},\
	\bibinfo {series and number} {STOC '11}\ (\bibinfo  {publisher} {ACM},\
	\bibinfo {address} {New York, NY, USA},\ \bibinfo {year} {2011})\ pp.\
	\bibinfo {pages} {333--342}\BibitemShut {NoStop}%
	\bibitem [{\citenamefont {Wang}\ \emph {et~al.}(2019)\citenamefont {Wang},
		\citenamefont {Qin}, \citenamefont {Ding}, \citenamefont {Chen},
		\citenamefont {Chen}, \citenamefont {You}, \citenamefont {He}, \citenamefont
		{Jiang}, \citenamefont {You}, \citenamefont {Wang}, \citenamefont
		{Schneider}, \citenamefont {Renema}, \citenamefont {H\"ofling}, \citenamefont
		{Lu},\ and\ \citenamefont {Pan}}]{Wang:2019}%
	\BibitemOpen
	\bibfield  {author} {\bibinfo {author} {\bibfnamefont {H.}~\bibnamefont
			{Wang}}, \bibinfo {author} {\bibfnamefont {J.}~\bibnamefont {Qin}}, \bibinfo
		{author} {\bibfnamefont {X.}~\bibnamefont {Ding}}, \bibinfo {author}
		{\bibfnamefont {M.-C.}\ \bibnamefont {Chen}}, \bibinfo {author}
		{\bibfnamefont {S.}~\bibnamefont {Chen}}, \bibinfo {author} {\bibfnamefont
			{X.}~\bibnamefont {You}}, \bibinfo {author} {\bibfnamefont {Y.-M.}\
			\bibnamefont {He}}, \bibinfo {author} {\bibfnamefont {X.}~\bibnamefont
			{Jiang}}, \bibinfo {author} {\bibfnamefont {L.}~\bibnamefont {You}}, \bibinfo
		{author} {\bibfnamefont {Z.}~\bibnamefont {Wang}}, \bibinfo {author}
		{\bibfnamefont {C.}~\bibnamefont {Schneider}}, \bibinfo {author}
		{\bibfnamefont {J.~J.}\ \bibnamefont {Renema}}, \bibinfo {author}
		{\bibfnamefont {S.}~\bibnamefont {H\"ofling}}, \bibinfo {author}
		{\bibfnamefont {C.-Y.}\ \bibnamefont {Lu}},\ and\ \bibinfo {author}
		{\bibfnamefont {J.-W.}\ \bibnamefont {Pan}},\ }\bibfield  {title} {\bibinfo
		{title} {Boson sampling with 20 input photons and a 60-mode interferometer in
			a $1{0}^{14}$-dimensional {H}ilbert space},\ }\href
	{https://doi.org/10.1103/PhysRevLett.123.250503} {\bibfield  {journal}
		{\bibinfo  {journal} {Phys. Rev. Lett.}\ }\textbf {\bibinfo {volume} {123}},\
		\bibinfo {pages} {250503} (\bibinfo {year} {2019})}\BibitemShut {NoStop}%
	\bibitem [{\citenamefont {Pleinert}\ \emph {et~al.}(2020)\citenamefont
		{Pleinert}, \citenamefont {von Zanthier},\ and\ \citenamefont
		{Lutz}}]{Pleinert:2020}%
	\BibitemOpen
	\bibfield  {author} {\bibinfo {author} {\bibfnamefont {M.-O.}\ \bibnamefont
			{Pleinert}}, \bibinfo {author} {\bibfnamefont {J.}~\bibnamefont {von
				Zanthier}},\ and\ \bibinfo {author} {\bibfnamefont {E.}~\bibnamefont
			{Lutz}},\ }\bibfield  {title} {\bibinfo {title} {Many-particle interference
			to test {B}orn's rule},\ }\href
	{https://doi.org/10.1103/PhysRevResearch.2.012051} {\bibfield  {journal}
		{\bibinfo  {journal} {Phys. Rev. Research}\ }\textbf {\bibinfo {volume}
			{2}},\ \bibinfo {pages} {012051(R)} (\bibinfo {year} {2020})}\BibitemShut
	{NoStop}%
	\bibitem [{Sup()}]{Supplemental}%
	\BibitemOpen
	\href@noop {} {}\bibinfo {note} {See Supplemental Material at ... for
		additional details regarding the experimental setup.}\BibitemShut {Stop}%
	\bibitem [{\citenamefont {Yabuki}(1986)}]{Yabuki:1986}%
	\BibitemOpen
	\bibfield  {author} {\bibinfo {author} {\bibfnamefont {H.}~\bibnamefont
			{Yabuki}},\ }\bibfield  {title} {\bibinfo {title} {Feynman path integrals in
			the {Y}oung double-slit experiment},\ }\href
	{https://doi.org/10.1007/BF00677704} {\bibfield  {journal} {\bibinfo
			{journal} {Int. J. Theor. Phys.}\ }\textbf {\bibinfo {volume} {25}},\
		\bibinfo {pages} {159} (\bibinfo {year} {1986})}\BibitemShut {NoStop}%
	\bibitem [{\citenamefont {Sawant}\ \emph {et~al.}(2014)\citenamefont {Sawant},
		\citenamefont {Samuel}, \citenamefont {Sinha}, \citenamefont {Sinha},\ and\
		\citenamefont {Sinha}}]{Sawant:2014}%
	\BibitemOpen
	\bibfield  {author} {\bibinfo {author} {\bibfnamefont {R.}~\bibnamefont
			{Sawant}}, \bibinfo {author} {\bibfnamefont {J.}~\bibnamefont {Samuel}},
		\bibinfo {author} {\bibfnamefont {A.}~\bibnamefont {Sinha}}, \bibinfo
		{author} {\bibfnamefont {S.}~\bibnamefont {Sinha}},\ and\ \bibinfo {author}
		{\bibfnamefont {U.}~\bibnamefont {Sinha}},\ }\bibfield  {title} {\bibinfo
		{title} {Nonclassical paths in quantum interference experiments},\ }\href
	{https://doi.org/10.1103/PhysRevLett.113.120406} {\bibfield  {journal}
		{\bibinfo  {journal} {Phys. Rev. Lett.}\ }\textbf {\bibinfo {volume} {113}},\
		\bibinfo {pages} {120406} (\bibinfo {year} {2014})}\BibitemShut {NoStop}%
	\bibitem [{\citenamefont {Maga{\~n}a-Loaiza}\ \emph {et~al.}(2016)\citenamefont
		{Maga{\~n}a-Loaiza}, \citenamefont {De~Leon}, \citenamefont {Mirhosseini},
		\citenamefont {Fickler}, \citenamefont {Safari}, \citenamefont {Mick},
		\citenamefont {McIntyre}, \citenamefont {Banzer}, \citenamefont {Rodenburg},
		\citenamefont {Leuchs},\ and\ \citenamefont {Boyd}}]{Magana-Loaiza:2016}%
	\BibitemOpen
	\bibfield  {author} {\bibinfo {author} {\bibfnamefont {O.~S.}\ \bibnamefont
			{Maga{\~n}a-Loaiza}}, \bibinfo {author} {\bibfnamefont {I.}~\bibnamefont
			{De~Leon}}, \bibinfo {author} {\bibfnamefont {M.}~\bibnamefont
			{Mirhosseini}}, \bibinfo {author} {\bibfnamefont {R.}~\bibnamefont
			{Fickler}}, \bibinfo {author} {\bibfnamefont {A.}~\bibnamefont {Safari}},
		\bibinfo {author} {\bibfnamefont {U.}~\bibnamefont {Mick}}, \bibinfo {author}
		{\bibfnamefont {B.}~\bibnamefont {McIntyre}}, \bibinfo {author}
		{\bibfnamefont {P.}~\bibnamefont {Banzer}}, \bibinfo {author} {\bibfnamefont
			{B.}~\bibnamefont {Rodenburg}}, \bibinfo {author} {\bibfnamefont
			{G.}~\bibnamefont {Leuchs}},\ and\ \bibinfo {author} {\bibfnamefont {R.~W.}\
			\bibnamefont {Boyd}},\ }\bibfield  {title} {\bibinfo {title} {Exotic looped
			trajectories of photons in three-slit interference},\ }\href
	{http://dx.doi.org/10.1038/ncomms13987} {\bibfield  {journal} {\bibinfo
			{journal} {Nat. Commun.}\ }\textbf {\bibinfo {volume} {7}},\ \bibinfo {pages}
		{13987} (\bibinfo {year} {2016})}\BibitemShut {NoStop}%
	\bibitem [{Note1()}]{Note1}%
	\BibitemOpen
	\bibinfo {note} {Since the SPADs are not number-resolving, multi-photon
		events at a single SPAD cannot be excluded. However, such events have a
		negligible effect on our results in the parameter regime of the
		experiment~\cite {Supplemental}.}\BibitemShut {Stop}%
	\bibitem [{Note2()}]{Note2}%
	\BibitemOpen
	\bibinfo {note} {The two-particle Hong-Ou-Mandel experiment corresponds to
		second-order interference, since it only involves two input modes~\cite
		{Hong:1987}.}\BibitemShut {Stop}%
	\bibitem [{Note3()}]{Note3}%
	\BibitemOpen
	\bibinfo {note} {Second- and third-order terms are a little smaller than
		expected, which suggests a slight off-center measurement.}\BibitemShut
	{Stop}%
	\bibitem [{Note4()}]{Note4}%
	\BibitemOpen
	\bibinfo {note} {Due to a difficult alignment, this result has been obtained
		with $100$ re-measurements of only the configuration $ABCD$ and its
		sub-configurations.}\BibitemShut {Stop}%
	\bibitem [{\citenamefont {Keil}\ \emph {et~al.}(2016)\citenamefont {Keil},
		\citenamefont {Kaufmann}, \citenamefont {Kauten}, \citenamefont {Gstir},
		\citenamefont {Dittel}, \citenamefont {Heilmann}, \citenamefont {Szameit},\
		and\ \citenamefont {Weihs}}]{Keil:2016}%
	\BibitemOpen
	\bibfield  {author} {\bibinfo {author} {\bibfnamefont {R.}~\bibnamefont
			{Keil}}, \bibinfo {author} {\bibfnamefont {T.}~\bibnamefont {Kaufmann}},
		\bibinfo {author} {\bibfnamefont {T.}~\bibnamefont {Kauten}}, \bibinfo
		{author} {\bibfnamefont {S.}~\bibnamefont {Gstir}}, \bibinfo {author}
		{\bibfnamefont {C.}~\bibnamefont {Dittel}}, \bibinfo {author} {\bibfnamefont
			{R.}~\bibnamefont {Heilmann}}, \bibinfo {author} {\bibfnamefont
			{A.}~\bibnamefont {Szameit}},\ and\ \bibinfo {author} {\bibfnamefont
			{G.}~\bibnamefont {Weihs}},\ }\bibfield  {title} {\bibinfo {title} {Hybrid
			waveguide-bulk multi-path interferometer with switchable amplitude and
			phase},\ }\href {https://doi.org/10.1063/1.4960204} {\bibfield  {journal}
		{\bibinfo  {journal} {APL Photonics}\ }\textbf {\bibinfo {volume} {1}},\
		\bibinfo {pages} {081302} (\bibinfo {year} {2016})},\ \Eprint
	{https://arxiv.org/abs/https://doi.org/10.1063/1.4960204}
	{https://doi.org/10.1063/1.4960204} \BibitemShut {NoStop}%
	\bibitem [{Note5()}]{Note5}%
	\BibitemOpen
	\bibinfo {note} {\protect \leavevmode {\protect The
			sensitivity quantifies the smallest change that can be detected in a
			measurement and depends on the measurement method. On the other hand, the
			precision is related to the measurement errors of the experimental
			setup.}}\BibitemShut {Stop}%
	\bibitem [{\citenamefont {Oppel}\ \emph {et~al.}(2014)\citenamefont {Oppel},
		\citenamefont {Wiegner}, \citenamefont {Agarwal},\ and\ \citenamefont {von
			Zanthier}}]{Oppel:2014}%
	\BibitemOpen
	\bibfield  {author} {\bibinfo {author} {\bibfnamefont {S.}~\bibnamefont
			{Oppel}}, \bibinfo {author} {\bibfnamefont {R.}~\bibnamefont {Wiegner}},
		\bibinfo {author} {\bibfnamefont {G.~S.}\ \bibnamefont {Agarwal}},\ and\
		\bibinfo {author} {\bibfnamefont {J.}~\bibnamefont {von Zanthier}},\
	}\bibfield  {title} {\bibinfo {title} {Directional superradiant emission from
			statistically independent incoherent nonclassical and classical sources},\
	}\href {https://doi.org/10.1103/PhysRevLett.113.263606} {\bibfield  {journal}
		{\bibinfo  {journal} {Phys. Rev. Lett.}\ }\textbf {\bibinfo {volume} {113}},\
		\bibinfo {pages} {263606} (\bibinfo {year} {2014})}\BibitemShut {NoStop}%
	\bibitem [{\citenamefont {Sabuncu}\ \emph {et~al.}(2007)\citenamefont
		{Sabuncu}, \citenamefont {Mi\ifmmode~\check{s}\else \v{s}\fi{}ta},
		\citenamefont {Fiur\'a\ifmmode~\check{s}\else \v{s}\fi{}ek}, \citenamefont
		{Filip}, \citenamefont {Leuchs},\ and\ \citenamefont
		{Andersen}}]{Sabuncu:2007}%
	\BibitemOpen
	\bibfield  {author} {\bibinfo {author} {\bibfnamefont {M.}~\bibnamefont
			{Sabuncu}}, \bibinfo {author} {\bibfnamefont {L.}~\bibnamefont
			{Mi\ifmmode~\check{s}\else \v{s}\fi{}ta}}, \bibinfo {author} {\bibfnamefont
			{J.}~\bibnamefont {Fiur\'a\ifmmode~\check{s}\else \v{s}\fi{}ek}}, \bibinfo
		{author} {\bibfnamefont {R.}~\bibnamefont {Filip}}, \bibinfo {author}
		{\bibfnamefont {G.}~\bibnamefont {Leuchs}},\ and\ \bibinfo {author}
		{\bibfnamefont {U.~L.}\ \bibnamefont {Andersen}},\ }\bibfield  {title}
	{\bibinfo {title} {Nonunity gain minimal-disturbance measurement},\ }\href
	{https://doi.org/10.1103/PhysRevA.76.032309} {\bibfield  {journal} {\bibinfo
			{journal} {Phys. Rev. A}\ }\textbf {\bibinfo {volume} {76}},\ \bibinfo
		{pages} {032309} (\bibinfo {year} {2007})}\BibitemShut {NoStop}%
	\bibitem [{\citenamefont {Lassen}\ \emph
		{et~al.}(2010{\natexlab{a}})\citenamefont {Lassen}, \citenamefont {Sabuncu},
		\citenamefont {Huck}, \citenamefont {Niset}, \citenamefont {Leuchs},
		\citenamefont {Cerf},\ and\ \citenamefont {Andersen}}]{Lassen:2010}%
	\BibitemOpen
	\bibfield  {author} {\bibinfo {author} {\bibfnamefont {M.}~\bibnamefont
			{Lassen}}, \bibinfo {author} {\bibfnamefont {M.}~\bibnamefont {Sabuncu}},
		\bibinfo {author} {\bibfnamefont {A.}~\bibnamefont {Huck}}, \bibinfo {author}
		{\bibfnamefont {J.}~\bibnamefont {Niset}}, \bibinfo {author} {\bibfnamefont
			{G.}~\bibnamefont {Leuchs}}, \bibinfo {author} {\bibfnamefont {N.~J.}\
			\bibnamefont {Cerf}},\ and\ \bibinfo {author} {\bibfnamefont {U.~L.}\
			\bibnamefont {Andersen}},\ }\bibfield  {title} {\bibinfo {title} {Quantum
			optical coherence can survive photon losses using a continuous-variable
			quantum erasure-correcting code},\ }\href
	{https://doi.org/10.1038/nphoton.2010.168} {\bibfield  {journal} {\bibinfo
			{journal} {Nature Photonics}\ }\textbf {\bibinfo {volume} {4}},\ \bibinfo
		{pages} {700} (\bibinfo {year} {2010}{\natexlab{a}})}\BibitemShut {NoStop}%
	\bibitem [{\citenamefont {Lassen}\ \emph
		{et~al.}(2010{\natexlab{b}})\citenamefont {Lassen}, \citenamefont {Madsen},
		\citenamefont {Sabuncu}, \citenamefont {Filip},\ and\ \citenamefont
		{Andersen}}]{Lassen:2010a}%
	\BibitemOpen
	\bibfield  {author} {\bibinfo {author} {\bibfnamefont {M.}~\bibnamefont
			{Lassen}}, \bibinfo {author} {\bibfnamefont {L.~S.}\ \bibnamefont {Madsen}},
		\bibinfo {author} {\bibfnamefont {M.}~\bibnamefont {Sabuncu}}, \bibinfo
		{author} {\bibfnamefont {R.}~\bibnamefont {Filip}},\ and\ \bibinfo {author}
		{\bibfnamefont {U.~L.}\ \bibnamefont {Andersen}},\ }\bibfield  {title}
	{\bibinfo {title} {Experimental demonstration of squeezed-state quantum
			averaging},\ }\href {https://doi.org/10.1103/PhysRevA.82.021801} {\bibfield
		{journal} {\bibinfo  {journal} {Phys. Rev. A}\ }\textbf {\bibinfo {volume}
			{82}},\ \bibinfo {pages} {021801(R)} (\bibinfo {year}
		{2010}{\natexlab{b}})}\BibitemShut {NoStop}%
\end{thebibliography}
\end{document}

\title{Supplemental Material - Testing higher-order quantum interference with many-particle states}

\author{Marc-Oliver Pleinert}
\affiliation{Institut f\"{u}r Optik, Information und Photonik, \\ Friedrich-Alexander-Universit\"{a}t Erlangen-N\"{u}rnberg (FAU), 91058 Erlangen, Germany}
\affiliation{Erlangen Graduate School in Advanced Optical Technologies (SAOT), Friedrich-Alexander-Universit\"{a}t Erlangen-N\"{u}rnberg (FAU), 91052 Erlangen, Germany}
\author{Alfredo Rueda}
\affiliation{Institut f\"{u}r Optik, Information und Photonik, \\ Friedrich-Alexander-Universit\"{a}t Erlangen-N\"{u}rnberg (FAU), 91058 Erlangen, Germany}
\affiliation{{\rm{currently at:}} Scantinel Photonics, Carl-Zeiss-Strasse 22, 73447 Oberkochen, Germany}
\author{Eric Lutz}
\affiliation{Institute for Theoretical Physics I, University of Stuttgart, D-70550 Stuttgart, Germany}
\author{Joachim von Zanthier}
\affiliation{Institut f\"{u}r Optik, Information und Photonik, \\ Friedrich-Alexander-Universit\"{a}t Erlangen-N\"{u}rnberg (FAU), 91058 Erlangen, Germany}
\affiliation{Erlangen Graduate School in Advanced Optical Technologies (SAOT), Friedrich-Alexander-Universit\"{a}t Erlangen-N\"{u}rnberg (FAU), 91052 Erlangen, Germany}

\maketitle

In this Supplemental Material, we present further details of the experiment. In particular, we examine the different aspects \textit{light preparation}, \textit{light manipulation}, and \textit{light measurement}. Within each of the three topics, we present the related setup and motivate our choice of parameters.
We also discuss possible errors that may arise leading to a deviation from the theoretical expectation and estimate their order.
Note that the impact of various experimental imperfections/errors onto the Sorkin parameter has also been discussed in the Supplemental Material of Ref.~\cite{Kauten:2017}.

\subsection{Light preparation: Laser and light propagation to the slit mask}

\begin{figure*}
	\centering \includegraphics[width=\columnwidth]{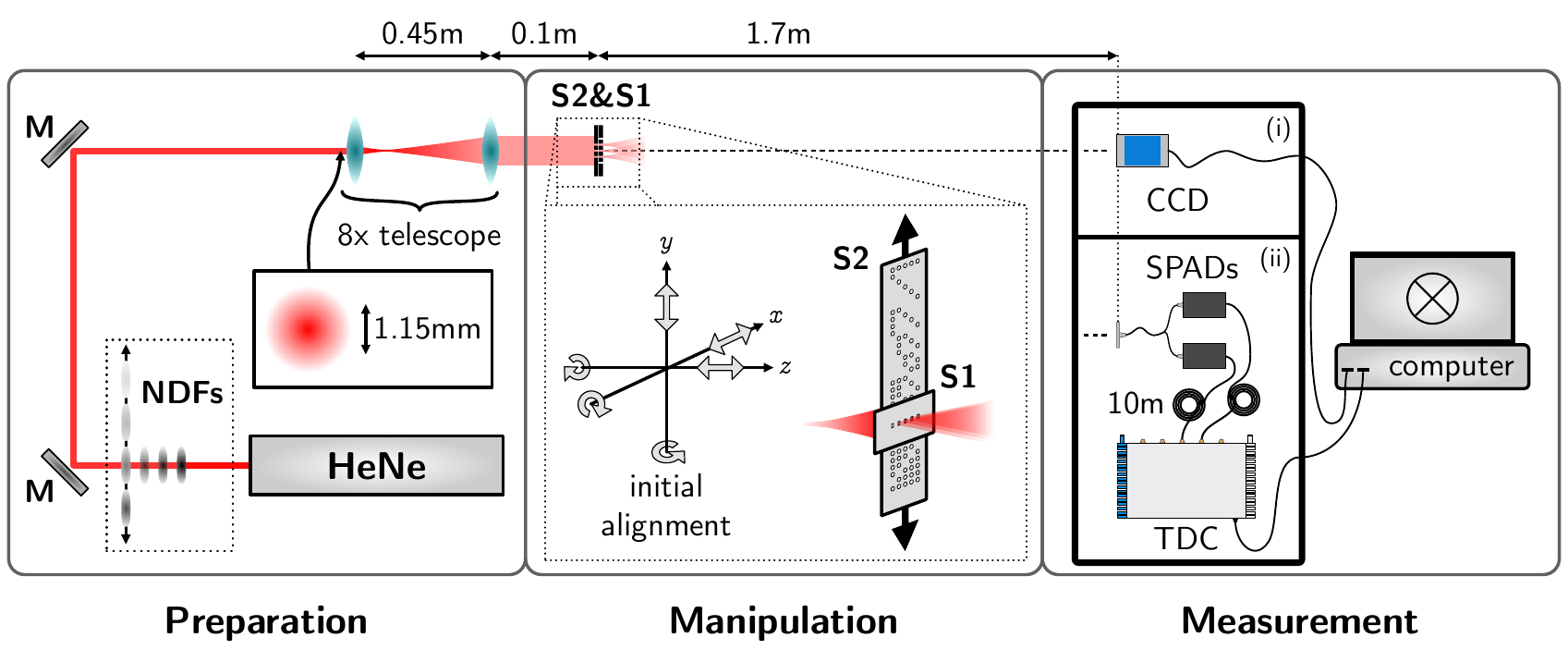}
	\caption{\label{fig:setup-detailed} Sketch of the experimental setup. HeNe: helium neon laser; NDFs: neutral density filters; M: mirror; S1, S2: slit masks; CCD: charge-coupled device for intensity measurements; SPADs: single-photon avalanche diodes for photon counting measurements; TDC: time-to-digital converter. For details, see the respective sections on preparation, manipulation, and measurement.}
\end{figure*}

In the experiment, we use a HeNe laser at $633\,\text{nm}$ (R-30995 from REO Inc.), with an output power of $17\,\text{mW}$, linear polarization $>500:1$, emitted essentially in the fundamental mode $\text{TEM}_{00}$ ($>99\%$). The laser is attenuated by various neutral density filters of different optical densities (OD), with a final flexible filter which can be switched between OD $0,0.6,1,2,3,\ldots$ to finetune the laser power. Behind the neutral density filters, two mirrors placed at right angles allow for precise alignment of the laser beam w.r.t. to the slit masks and the detection unit.
About $55\,\text{cm}$ before the slit masks, the laser beam is expanded by an $8\times$ telescope consisting of two lenses to ensure homogeneous illumination of all slits. 
We chose the beam to be slightly converging after the telescope, with the focal point being located $1.7\, \text{m}$ after the slit masks, where we put the detector system, the latter thus being located in the Fourier plane of the slit masks (see Fig.~\ref{fig:setup-detailed}). 

The coherence length of the laser is $\gtrsim 0.5\,\text{m}$, i.e., by far exceeding the path difference between any two slits of the slit masks, thus ensuring coherence of the diffracted light.

The high degree of polarization of the laser is principally not needed for the experiment, since 

even if the diffraction by the slit masks would be polarization sensitive, this would affect all measurements equally and thus would cancel out in the final calculation of the interference terms and the Sorkin parameter.

After warm-up of the laser, the relative intensity fluctuations are $<10^{-4}$ over the period of a typical measurement set. 
This has been determined by recording the laser power of the slit configuration ABCDE repeatedly during the same period. 
The intensity fluctuations of the laser are one of the main contributions to the uncertainty budget of the experiment. They could be improved in the future by actively stabilizing the laser power.

The spot size of the Gaussian beam before the telescope has been measured to $\approx 1.15 \, \text{mm}$; after
the telescope, the beam is expanded roughly to $9\,\text{mm}$ at the slit masks. 
Consequently, within the central $2\,\text{mm}$ of the beam corresponding to the extension of the five slits ABCDE ($4d+a$ in $x$-direction, see Fig.~1(b) in the main text), the variation of the laser intensity is smaller than $1\%$. 
Note that any residual variation in the $y$-direction influences all slits equally and therefore would not affect the interference terms and the Sorkin parameter.

\subsection{Light manipulation: Slit masks and their alignment}

The slit mask S1 has been produced with high precision using laser lithography and electron-beam physical vapor deposition by the Micro- \& Nanostructuring Unit of the Max Planck Institute for the Science of Light. The slit mask S2 - merely used for blocking the slits of S1 and with slit sizes larger than those of S1 (for a quantitative discussion see below) - has been manufactured by drilling holes in an aluminum plate.

In the experiment, the initial position and orientation of the slit masks can be controlled in various ways.
The slit mask S1 is mounted on an $xyz$-axis translation stage and can further be rotated around all three axes $x$, $y$, and $z$ (see also Fig.~\ref{fig:setup-detailed}).
This way, S1 can be precisely positioned to ensure a homogeneous illumination of all slits by the laser, with the slits being accurately aligned along the $x$-axis.
The blocking slit mask S2 is mounted on a translation stage, electronically movable along the $y$-direction with a stepper motor (Thorlabs LTS150). 
The translation stage itself is aligned manually such that S2 lies in the $xy$-plane, with the slits of S2 being parallel to the slits of S1.
The distance between the two slit masks, $L_{21}$, is less than $1\, \text{mm}$.
With this distance, the S2-to-S1 Fresnel number to $F_{21}=r^2/(L_{21} \lambda) > 252 \gg 1$;
the propagation from S2 to S1 can thus be approximated by geometrical optics and the diffraction by S2 be neglected at S1 in the experiment.
In this way, the slit mask S2 can be considered as a `mere' blocking device for the slits of S1 so that only the unblocked slits of S1 determine the interference pattern.
For all measurements, the mask S1 is fixed,
while the blocking slit mask S2 is moved in front of S1 via the translation stage. 
The whole measurement process, i.e., switching between different slit configurations and the measurement itself (see Sect.~\ref{sec:measurement} below), is automatized to improve precision.

In principle, the experiment could also be conducted with a single slit mask S2 with a large spacing $\Delta$ and directly scanning all different slit configurations of S2.
However, using two slit masks drastically reduces alignment uncertainties and suppresses errors resulting from imperfections of the masks.
With two slit masks, we can scan all required slit configurations by moving merely the blocking slit mask S2, while keeping the remaining components fixed, in particular the slit mask S1, the laser system, and the detection unit. This way, alignment errors induced by switching between configurations are substantially reduced. 
Moreover, using two slit masks, any imperfections of the slits of the mask S1 (e.g., slit A) affect in principle all configurations that contain this slit (e.g., A,AB,AC,...,ABCDE). Yet, in interference orders that are expected to be zero, these imperfections cancel due to the symmetric appearance of A, whereas in nonzero orders, those imperfections are typically much smaller than the signal itself.
Using only a single mask scanned between the required slit configurations, different imperfections will occur in different configurations, e.g., slit A in configurations AB and AC will lead to distinct deviations. These distinct deviations do not cancel and would thus lead to a systematic offset in the nonzero orders. 

\begin{figure*}
	\centering \includegraphics[width=\columnwidth]{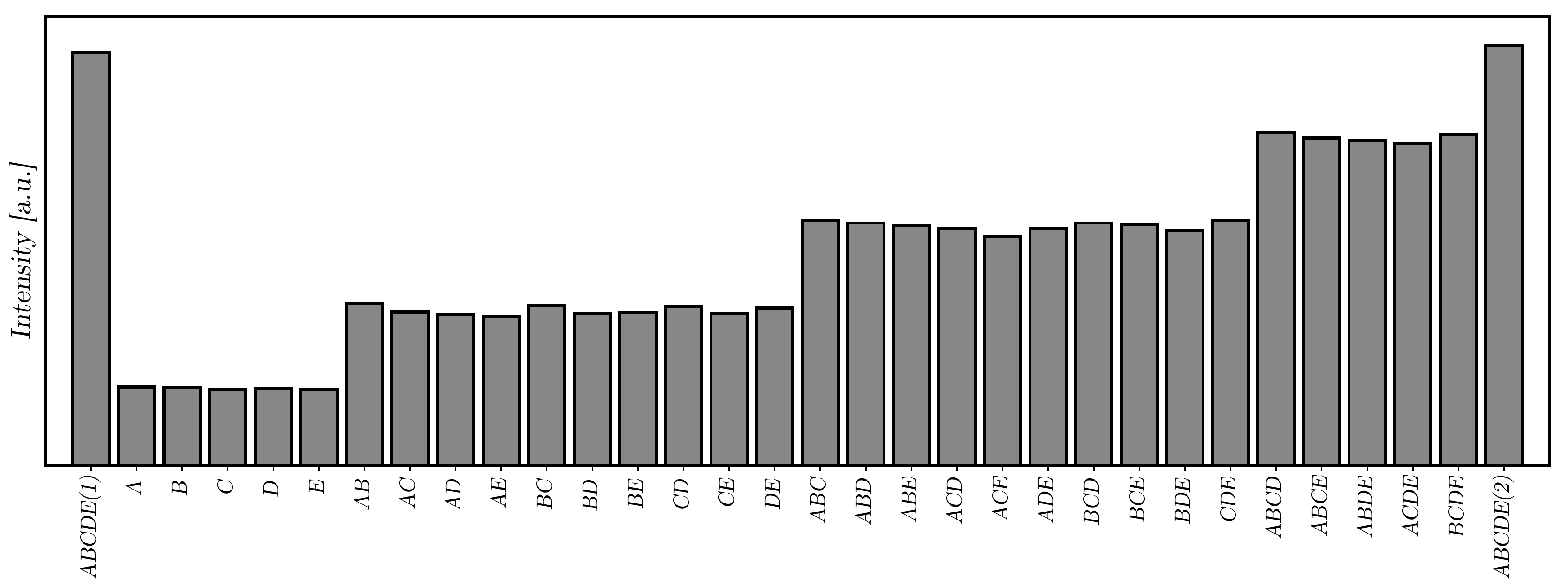}
	\caption{\label{fig:config-intensities} Experimental results of the intensity histogram for the $32$ slit configurations of the setup (background corrected).}
\end{figure*}

The dimensions of the slits in S1 have been chosen to $d > a,b > \lambda$. In this way, we obtain rapid oscillations of the interference pattern along the $x$-direction while diffraction effects induced by a single slit are kept small.
In the end, we opted for $d=500\,\mu\text{m}$, $b=200\,\mu\text{m}$, and $a=25\,\mu\text{m}$, with the extension of the slits in the $y$-direction (where diffraction is irrelevant) being larger than in the $x$-direction to increase the amount of light passing through the slit system. The percentage of light passing through a single slit of S1 is $\approx 1.2\permil$ of the incident light for the Gaussian beam used in the experiment; for configurations with more slits it is accordingly higher.
The experimentally measured relative intensities for the different slit configuration are shown in Fig.~\ref{fig:config-intensities}. The intensities are averaged over the cropped images (see below for details about the evaluation process) and clearly show a linear dependency with the number of slits.

The diameter $2r$ of the holes in the blocking mask S2 is greater than $b$ and smaller than $d$ to cover accurately a single slit of S1 and simplify the alignment.
In the experiment, we used two different slit masks S2 with two distinct diameters $2r=300\,\mu\text{m}$ and $2r=400\,\mu\text{m}$, respectively. 
In the end, accurate relative alignment of the two masks S1 and S2 was easier to achieve with the greater slit diameter $2r=400\,\mu\text{m}$.
The distance between the different slit configurations $\Delta$ has to be greater than $r$, but, in principle, can be arbitrarily large. 
However, the larger $\Delta$, the greater the possibility of relative misalignment of the two slit masks when switching from one configuration to the next one. 
Thus, we finally chose $\Delta=1000\,\mu\text{m}$.

\subsection{Light measurement: Propagation of the light behind the slit masks and detection systems}\label{sec:measurement}

The chosen distance $L=1.7\,\text{m}$ between S1 and the detection unit is on the one hand far enough such that the CCD resolves precisely the interference pattern and, on the other hand, close enough such that a sufficient amount of light reaches the detector.
The Fresnel number in the detection plane is $F_1 = ab/(L\lambda) \approx 5\cdot10^{-3} \ll 1$, i.e., the detection is conducted in the Fraunhofer regime.

\subsubsection{CCD}

\paragraph{Parameters.}

The diffracted light behind the slit masks is recorded with the CCD \textit{pco.pixelfly} from PCO AG. 
It has a dynamical range of $14$ bits and consists of $1392 \times 1040$ pixels of size $(d_{pixel})^2=(6.45 \, \mu m)^2$. 
At $L=1.7\,\text{m}$, the CCD covers about $8\pi$ (in terms of the optical phase $\delta_p$, see below) of the resulting interference pattern in the $x$-direction.
To exclude overflow errors of the CCD towards the image edges, we cut the CCD images and used only the central $\approx 6\pi$ for the evaluation.
In this way we cover three periods of the interference pattern, i.e., within $\delta \in [-3\pi,3\pi]$.
In $y$-direction, the CCD covers only $b\sin (\theta)/\lambda \in [-0.62\pi,0.62\pi]$ of the diffraction pattern.
To exclude edge effects, we again cut the images in this direction and ultimately used the portion $[-0.5\pi,0.5\pi]$.
In total, we thus used $\approx 1000 \times 850  \approx 8.5 \cdot 10^5$ pixels of the CCD as independent detectors.

At $633\,\text{nm}$, the quantum efficiency of the CCD is $\approx 40\%$.
Hence, the real signal of the diffraction patterns is higher than the measured one. 
However, as long as this affects all measurements equally, the effect cancels in the properly normalized interference orders as it influences numerator and denominator in the same way.

The integration time of the CCD $t_i=2\,\text{ms}$ has been chosen to fully cover the dynamical range of the CCD for the five-slit configuration ABCDE (resulting in a maximum amount of light recorded by the CCD).
An increased integration time would lead to saturation, while a decreased integration time would lead to less precision.
To average over short-term fluctuations, while, at the same time, avoiding long-term fluctuations (and to keep the entire measurement time feasible), we chose to use $250$ images per configuration.  
Moreover, test runs with $100$ images per configuration yielded worse results since short term fluctuations were not yet sufficiently averaged out, while test runs with $1000$ images per configuration did not improve the results significantly. 

\paragraph{Data evaluation process.}

Due to the 1D-geometry of the slit configurations, the modulation of the interference patterns only occurs along horizontal lines of the CCD (i.e., along the $x$-direction).
Since each pixel $p$ of the CCD can be identified as an individual detector, each horizontal line of the CCD constitutes an independent measurement of the 1D single-particle interference pattern.
Photons emanating from one slit of the mask and hitting the $p$-th pixel within a line of the CCD accumulate the phase (with respect to a photon emanating from an adjacent slit) 
\begin{equation}
\delta_p = \frac{2\pi}{\lambda} d \frac{p \cdot d_{pixel}}{L} \, .
\end{equation}
Here, $\lambda$ is the HeNe laser wavelength, $d$ the inter-slit spacing, $L$ the distance between the slits of the mask and the CCD, and $d_{pixel}$ the size of an individual pixel. Note that only relative phases are relevant for the interference pattern, which is why the enumeration $p$ has to be consistent, but, in principle, can be arbitrary.

\begin{figure*}
	\centering \includegraphics[width=\columnwidth]{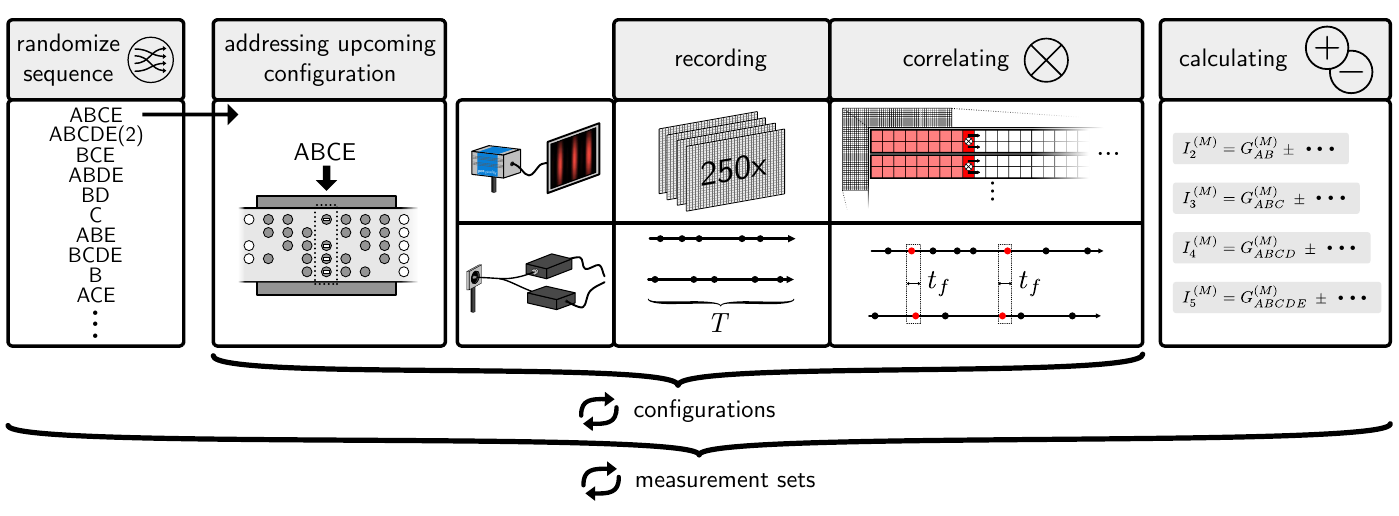}
	\caption{\label{fig:meas-process} Fully automated measurement process and calculation of the interference hierarchy. For each measurement set, the sequence of slit configurations is first randomly chosen. According to the drawn order, a single configuration is addressed via the translation stage. Then, either $250$ images are recorded with the CCD, or particle counts are registered with two SPADs. The resulting data is then correlated yielding the correlation functions $G^{(M)}_X$ per configuration $X$. After all different configurations have been measured, the resulting correlation functions are corrected by the measured background and the interference orders are calculated according to their definition.}
\end{figure*}

For two-particle correlations, we correlate the intensity of one pixel from a given horizontal line of the CCD and correlate it with the intensity of a pixel of a neighboring horizontal line in order to exclude self-correlations (correlations of the same detector (pixel) with itself).
For the second-order correlation functions, $G^{(2)}(\delta_1,\delta_2)$, the optical phase $\delta_1$ ($\delta_2$) is then always associated to a pixel of the first (second) line in such a measurement. 
This way, we always detect correlations between independent detectors (different pixels), even in the auto-correlation scheme ($\delta=\delta_1=\delta_2$). The latter is also sketched under `correlating' in the first line of Fig.~\ref{fig:meas-process}.

The interference orders are then calculated from the measured correlation functions via their definitions as given in the main text.
The values and uncertainties in the intensity regime are the arithmetic mean and the standard deviation of all different 1D-measurements.

\subsubsection{SPADs}

In the photon-counting regime, the measurement is conducted via a multi-mode fiber beam splitter connected to two SPADs (PDM \$PD-050-CTB-FC from MPD); the SPADs measure individual photons which are then fed into a time-to-digital converter (TDC) (quTAG from qutools GmbH). The photon time-tags of the TDC are then correlated by a computer program (see below).
Since we are interested in the auto-correlation scheme ($\delta=\delta_1=\delta_2$, see main text), we only measure photons stemming from a single mode. Therefore, we placed an additional TEM$_{00}$ single-mode fiber in front of the multi-mode fiber beam splitter. Light emanating from the slit masks is coupled into the single-mode fiber by use of a lens (L) for simplification.  In front of the lens, we use a pinhole (P) of radius $10 \,\mu\text{m}$ to collect a spatially confined part of the interference pattern.
The whole collection apparatus consisting of pinhole, lens, and single-mode fiber is mounted on an $xy$-translation stage. As a result, we are able to scan the interference pattern and search for the central maximum $\delta=0$, where the single- and two-particle hierarchy differ the most.

\paragraph{Parameters.}

In the experiment, we use SPADs (PDM Series of Micro Photon Devices) with a quantum efficiency of $\approx 40\%$ at $633\,\text{nm}$. As outlined above, a quantum efficiency $< 100\%$ does not influence the normalized interference orders.
The dead time of the SPADs is $77\,\text{ps}$ leading to a non-linear response of the detector for high count rates. 
Such non-linearities could in principle bias the result, as they have a stronger effect on the five-slit pattern than on the single-slit ones.
To avoid this bias, we adjusted in the experiment the laser intensity by use of the neutral density filters (see above) such that the count rate correspond to $< 10^5\,\text{Hz}$ for all slit configurations. In this case, any bias due to detector non-linearities is $\leq 10^{-3}$.

The two SPADs are connected with two $10\,\text{m}$ cables with the TDC.
The time resolution of the whole detection system is $\approx 90\,\text{ps}$. 
The total time of measuring per configuration is $T=120\,\text{s}$. 
This choice of $T$ yields sufficient statistics for single- and two-particle events, and is also a good choice for the above trade-off between short-term and long-term fluctuations (as discussed for the CCD). 
In test runs, the correlations have been evaluated with three different values of $t_f=100\,\text{ps},1\,\text{ns},10\,\text{ns}$. 
With $t_f=100\,\text{ps}$, however, insufficient coincidences have been collected, while with $t_f=10\,\text{ns}$, the two-particle interference hierarchy already reduced towards the single-particle one, due to the increased incoherence of the two collected photons ($10\,\text{ns}$ correspond to $3\,\text{m}$, which exceeds the coherence length of the laser).
$t_f=1\,\text{ns}$ can thus be considered as a sweet spot for the current experiment.
We note that the used SPADs are not number-resolving. 
Events, in which two or more particles are registered at one of the SPADs can thus not be differentiated from single-particle events. However, such events only have a minor effect on the interference orders: The ratio between the measured single-photon events and the measured two-photon events is on the order of $10^{-3}-10^{-4}$ (depending on the slit configuration). Using a $50:50$ beam splitter, where an incident two-photon state splits with a probability of $0.5$, the actual number of two-photon events is thus two times higher than the measured one. Hence, we actually over-count the single-particle events by the number of two-particle events and under-count the two-particle events by a factor of $2$. 
A correction of the final result via a corresponding multiplication/addition is straightforward but does not change the results of the \textit{normalized }two-particle interference orders: Correcting the measured two-photon correlations implies multiplying numerator and denominator of the normalized interference orders equally, which in the end thus cancels. 
Only the single-particle results would be affected by a factor $10^{-3}-10^{-4}$ what is below our final precision.
Going on to three-photon events, they also lead to an error in the two-particle investigations. However, these events are even more unlikely (assuming a Poissonian statistic, $\approx 10^{-3}-10^{-4}$ w.r.t. the two-particle events and thus $\approx 10^{-6}-10^{-8}$ w.r.t. the single-particle events). 
These events thus yield deviations which are again below our precision (e.g., of the order $10^{-3}-10^{-4}$ for the two-particle correlations).
In the future, such errors could be excluded by upgrading the detection system to number-resolving SPADs.

\paragraph{Data evaluation process.}

Out of all events recorded by the two SPADs and the TDC, we differentiate between the single- and two-particle hierarchy via sorting:
For two-particle correlations, we count the number of coincident events $N^{(2)} \propto P_\alpha^{(2)} $, i.e., whenever two events (one from SPAD 1 and one from SPAD 2) fall within the correlation time of $1\,\text{ns}$. The remaining counts $N^{(1)} \propto P_\alpha^{(1)}$, i.e., when only one of the two SPADs registered a photon, are single-particle events and yield the single-particle interference hierarchy; the recording and correlating in the photon-counting regime are also sketched in the second line of Fig.~\ref{fig:meas-process}.
The interference orders are then calculated via their definitions as given in the main text. 
Here, each measurement is associated with a Poissonian photocount error, which is then error-propagated to the interference orders. Note that in Fig.~3 of the main text, the resulting error is enlarged by a factor of 100 to make it more visible.